\documentclass[fleqn,usenatbib]{mnras}

\usepackage[T1]{fontenc}

\DeclareRobustCommand{\VAN}[3]{#2}
\let\VANthebibliography\thebibliography
\def\thebibliography{\DeclareRobustCommand{\VAN}[3]{##3}\VANthebibliography}

\usepackage{graphicx}	
\usepackage{amsmath}	
\usepackage{amssymb}
\usepackage{mathrsfs}
\usepackage{txfonts}
\usepackage[nopatch]{microtype}
\usepackage{booktabs}

\usepackage{tabularx}
\usepackage[skip=2pt]{caption}
\usepackage{cite}
\usepackage{orcidlink}

\usepackage{bm}
\usepackage{array}
\usepackage{relsize}
\usepackage{acro}
\usepackage{multicol}
\usepackage{multirow}
\usepackage{xfrac}
\usepackage{lipsum}
\usepackage{cleveref}

\crefname{figure}{Figure}{Figures}
\Crefname{figure}{Figure}{Figures}
\crefname{table}{Table}{Tables}
\Crefname{table}{Table}{Tables}

\usepackage{hyperref}

\DeclareAcronym{agn}{
  short = AGN ,
  short-plural-form = AGN ,
  long  = active galactic nucleus ,
  long-plural-form = active galactic nuclei ,
  tag = abbrev
}

\DeclareAcronym{smbh}{
  short = SMBH ,
  long  = supermassive black hole ,
  tag = abbrev
}

\DeclareAcronym{mq}{
  short = MQ ,
  long  = massive quiescent ,
  tag = abbrev
}

\DeclareAcronym{mqg}{
  short = MQG ,
  long  = massive quiescent galaxy ,
  long-plural-form = massive quiescent galaxies ,
  tag = abbrev
}

\DeclareAcronym{rq}{
  short = RQ ,
  long  = recent quiescent ,
  tag = abbrev
}

\DeclareAcronym{rqg}{
  short = RQG ,
  long  = recently quenched galaxy ,
  long-plural-form = recently quenched galaxies ,
  tag = abbrev
}

\DeclareAcronym{lq}{
  short = LQ ,
  long  = long quiescent ,
  tag = abbrev
}

\DeclareAcronym{lqg}{
  short = LQG ,
  long  = long quiescent galaxy ,
  long-plural-form = long quiescent galaxies ,
  tag = abbrev
}

\DeclareAcronym{coda}{
  short = CoDaIII ,
  long  = Cosmic Dawn III ,
  tag = abbrev
}

\DeclareAcronym{illtng}{
  short = TNG ,
  long  = Illustris: The Next Generation ,
  tag = abbrev
}

\DeclareAcronym{cpc}{
  short = cpc ,
  short-plural =  ,
  long  = comoving parsec ,
  tag = abbrev
}

\DeclareAcronym{ckpc}{
  short = ckpc ,
  short-plural =  ,
  long  = comoving kiloparsec ,
  tag = abbrev
}

\DeclareAcronym{cMpc}{
  short = cMpc ,
  short-plural =  ,
  long  = comoving megaparsec ,
  tag = abbrev
}

\DeclareAcronym{cGpc}{
  short = cGpc ,
  short-plural =  ,
  long  = comoving gigaparsec ,
  tag = abbrev
}

\DeclareAcronym{ppc}{
  short = ppc ,
  short-plural =  ,
  long  = physical parsec ,
  tag = abbrev
}

\DeclareAcronym{pkpc}{
  short = pkpc ,
  short-plural =  ,
  long  = physical kiloparsec ,
  tag = abbrev
}

\DeclareAcronym{pMpc}{
  short = pMpc ,
  short-plural =  ,
  long  = physical megaparsec ,
  tag = abbrev
}

\DeclareAcronym{pGpc}{
  short = pGpc ,
  short-plural =  ,
  long  = physical gigaparsec ,
  tag = abbrev
}

\DeclareAcronym{lt}{
  short = LT ,
  long  = logarithmic transformation ,
  tag = abbrev
}

\DeclareAcronym{gqt}{
  short = GQT ,
  long  = Gaussian quantile transformation ,
  tag = abbrev
}

\DeclareAcronym{cdf}{
  short = CDF ,
  long  = cumulative distribution function ,
  tag = abbrev
}

\DeclareAcronym{imf}{
  short = IMF ,
  long  = initial mass function ,
  tag = abbrev
}

\DeclareAcronym{sfh}{
  short = SFH ,
  long  = star formation history ,
  long-plural-form = star formation histories ,
  tag = abbrev
}

\DeclareAcronym{zh}{
  short = ZH ,
  long  = metallicity history ,
  long-plural-form = metallicity histories ,
  tag = abbrev
}

\DeclareAcronym{bgs}{
  short = BGS ,
  long  = Bright Galaxy Survey ,
  tag = abbrev
}

\DeclareAcronym{grispy}{
  short = GriSPy ,
  long  = Grid Search In Python ,
  tag = abbrev
}

\DeclareAcronym{disperse}{
  short = DisPerSE ,
  long  = Discrete Persistent Structure Extractor ,
  tag = abbrev
}

\DeclareAcronym{pca}{
  short = PCA ,
  long  = principal component analysis ,
  long-plural-form = principal component analyses ,
  tag = abbrev
}

\DeclareAcronym{pc}{
  short = PC ,
  long  = principal component ,
  tag = abbrev
}

\DeclareAcronym{lrd}{
  short = LRD ,
  long  = little red dot ,
  tag = abbrev
}

\DeclareAcronym{gmm}{
  short = GMM ,
  long  = Gaussian mixture model ,
  tag = abbrev
}

\DeclareAcronym{em}{
  short = EM ,
  long  = expectation-maximisation ,
  tag = abbrev
}

\DeclareAcronym{ghc}{
  short = GHC ,
  long  = galaxy-halo connection ,
  tag = abbrev
}

\DeclareAcronym{shmr}{
  short = SHMR ,
  long  = stellar-halo mass relation ,
  tag = abbrev
}

\DeclareAcronym{mzr}{
  short = MZR ,
  long  = mass-metallicity relation ,
  tag = abbrev
}

\DeclareAcronym{mhzr}{
  short = HMZR ,
  long  = halo mass-metallicity relation ,
  tag = abbrev
}

\DeclareAcronym{ssp}{
  short = SSP ,
  long  = simple stellar population ,
  tag = abbrev
}

\DeclareAcronym{srcc}{
  short = $r_S$ ,
  long  = Spearman's rank correlation coefficient ,
  tag = abbrev
}

\DeclareAcronym{mwa}{
  short = MWA ,
  long  = mass-weighted age ,
  tag = abbrev
}

\DeclareAcronym{zwa}{
  short = ZWA ,
  long  = metallicity-weighted age ,
  tag = abbrev
}

\DeclareAcronym{mpb}{
  short = MPB ,
  long  = main progenitor branch ,
  long-plural-form = main progenitor branches ,
  tag = abbrev
}

\DeclareAcronym{spl}{
  short = SPL ,
  long  = secondary progenitor list ,
  tag = abbrev
}

\DeclareAcronym{rfr}{
  short = RFR ,
  long  = random forest regressor ,
  tag = abbrev
}

\DeclareAcronym{ert}{
  short = ERT ,
  long  = extremely randomised tree ,
  tag = abbrev
}

\DeclareAcronym{fof}{
  short = FoF ,
  long  = Friends-of-Friends ,
  tag = abbrev
}

\DeclareAcronym{relu}{
  short = ReLU ,
  long  = rectified linear unit ,
  tag = abbrev
}

\DeclareAcronym{lrelu}{
  short = L-ReLU ,
  long  = leaky rectified linear unit ,
  tag = abbrev
}

\DeclareAcronym{elu}{
  short = ELU ,
  long  = exponential linear unit ,
  tag = abbrev
}

\DeclareAcronym{fsps}{
  short = FSPS ,
  long  = Flexible Stellar Population Synthesis ,
  tag = abbrev
}

\DeclareAcronym{UM}{
  short = UM ,
  long  = UniverseMachine ,
  tag = abbrev
}

\DeclareAcronym{desi}{
  short = DESI ,
  long  = Dark Energy Spectroscopic Instrument ,
  tag = abbrev
}

\DeclareAcronym{sam}{
  short = SAM ,
  long  = semi-analytic model ,
  tag = abbrev
}

\defcitealias{Chittenden}{CT23}
\defcitealias{Behera}{BTC24}

\title[The \textsc{Aurora} Black Hole Seeding Model]{Introducing \textsc{Aurora}: A New Black Hole Seeding Model With Explicit Environmental Coupling}

\author[H. G. Chittenden et al.]{
Harry George Chittenden$^{1,2}$\thanks{E-mail: hchittenden@swin.edu.au}\textsuperscript{\orcidlink{0000-0001-5856-8713}},
Darren Croton$^{1,3}$\textsuperscript{\orcidlink{0000-0002-5009-512X}},
Karl Glazebrook$^{1,2}$\textsuperscript{\orcidlink{0000-0002-3254-9044}} and \newauthor \ Robert Adriel Mostoghiu Paun$^{1,3}$\textsuperscript{\orcidlink{0000-0002-5425-0641}}
\\
$^{1}$Centre for Astrophysics and Supercomputing [CAS], Swinburne University of Technology, P.O. Box 218, Hawthorn VIC 3122, Melbourne, Australia\\
$^{2}$JWST Australia Data Centre [JADC], Swinburne Advanced Manufacturing and Design Centre [AMDC], John Street, Hawthorn VIC 3122, Australia\\
$^{3}$ARC Centre of Excellence for Dark Matter Particle Physics [CDM], Melbourne, Australia
}

\date{Accepted XXX. Received YYY; in original form ZZZ}

\pubyear{2026}

\begin{document}
\label{firstpage}
\pagerange{\pageref{firstpage}--\pageref{lastpage}}
\maketitle

\begin{abstract}
We present \textsc{Aurora}, a black hole seeding module for the \textsc{Mimic} semi-analytic framework, in which the seed mass of each newly resolved halo is set by its environment. Motivated by evidence that early quiescent galaxies and overmassive black holes assemble in dense environments, \textsc{Aurora} replaces the fixed seed masses of conventional prescriptions with a continuous spectrum, coupling the seed mass to two environmental parameters, the local overdensity and the radial skewness of the surrounding haloes, and to the halo mass at seeding. Requiring only halo masses and positions, it applies to dark-matter-only simulations. Calibrated against \textsc{Sage} on the \textsc{Millennium} merger trees, this prescription leaves a lasting signal. Massive black holes are assembled from the seeds of many progenitors, seeds supplying 29 per cent of all black hole mass formed by $z=0$. A seed-dominated growth pathway, distinct from the accretion-dominated majority, emerges among galaxies assembled rapidly in dense, asymmetric environments. Environment-dependent seeding raises the quiescent fraction of seeded galaxies in massive haloes at $z \simeq 3.1$ by up to 70 per cent, with the excess concentrated in overdense environments, and a deficit in voids. Scrambling the seeding environments while preserving the mass spectrum erases this gradient. The two environmental couplings imprint separable signatures, broadening the scatter of the galaxy-black hole mass relation while preserving the stellar-halo mass relation and other \textsc{Sage} summary statistics. We discuss applying \textsc{Aurora} at gigaparsec scales and higher resolution, to understand the origins of massive black hole systems in contemporary deep surveys.
\end{abstract}


\section{Introduction}
\label{sec:intro}

Observations with JWST have transformed the census of the earliest \acp{smbh}. Spectroscopy across the first billion years of cosmic history has revealed an abundant population of faint \acp{agn} \citep{Harikane2023, Taylor2025, Juodzbalis2026}, most strikingly the compact and ubiquitous \acp{lrd} \citep{Greene, Maiolino, Akins2025, Kocevski2025}, whose interpretation, as \ac{agn}-dominated, stellar-dominated or composite, remains open \citep{PerezGonzalez2024, Inayoshi2025b, deGraaff2026, Inayoshi2025, Weibel2}. A significant fraction of these systems appear overmassive relative to their hosts, with black hole-to-stellar mass ratios of $M_\text{BH}/M_\ast \sim 10^{-2}-10^{-1}$ at $z \sim 4-11$ \citep{Pacucci, Maiolino, Ubler, Ma2026}, one to two orders of magnitude above the local relation, and higher still in individual cases. These deep, narrow surveys are complemented by the wide area facilities now coming online, which trade depth for the volume needed to capture the rarest systems. For example, \textit{Euclid} is predicted to yield over a hundred quasars at $7.0 < z < 7.5$ from its Wide Survey, and $\sim\!25$ beyond $z = 7.5$ \citep{Barnett2019, EuclidCollaboration2022}. Its first $1.5$ years of data have begun to bear that out, with $31$ quasars at $6.6 < z < 7.8$, twelve of which lie at $z \geq 7$ \citep{EuclidCollaboration2026}. The \textit{Roman} High Latitude Spectroscopic Survey will map $2000\,\text{deg}^2$ to greater depth \citep{Wang2022}, building on the ground based quasar samples which first established the existence of black holes above $10^{9}\,\text{M}_\odot$ within the first gigayear of the Big Bang \citep{Fan2023}. At the opposite extreme, wide spectroscopic surveys constrain the low mass end of the occupation distribution, where the seeding signal is least diluted \citep{Pucha2025}. Together, these discoveries impose joint constraints on when, where, and at what mass the first black holes formed, and on how rapidly they subsequently grew.

The same surveys have exposed a parallel tension in the galaxies themselves. \Acp{mqg} are now confirmed spectroscopically well beyond $z \simeq 4$ \citep{Carnallnature, Valentino2023, deGraaff, Nanayakkara, Weibel, Ji2026, Zhang2026}, in numbers that most contemporary models underpredict by an order of magnitude or more \citep{Lagos, Chittenden3}, and in several cases within conspicuously overdense structures \citep{Tanaka2023, Ito}. Here too the wide surveys are beginning to contribute, with \textit{Euclid} identifying massive galaxy candidates towards the end of reionisation across its first quick data release \citep{EuclidCollaboration2025a} and isolating extremely red high redshift sources in its deep fields \citep{Signor2024}. The simulations that do produce them in comparable abundance achieve this through markedly different routes \citep{Xie, Kimmig, RemusKimmig, Seeyave2025, Szpila, ChandroGomez1, ChandroGomez2, Chon}, such that matching the census does not by itself identify the mechanism. Since quenching at these epochs is largely dependent on \ac{agn} feedback, and the strength of said feedback depends on how massive the black hole became and how early, the abundance of early quiescent galaxies and the masses of the earliest black holes are two aspects of a single problem.

The formation channels proposed for these objects span roughly four decades in initial mass (for reviews, see \citealt{Inayoshi2020, Regan2024}): light seeds of $\sim\!10^{2}-10^{3}\,M_\odot$, the remnants of metal-poor Population III stars; heavy seeds of $\sim\!10^{4}-10^{6}\,M_\odot$, formed by direct collapse in atomic cooling haloes where fragmentation is suppressed; and intermediate channels through runaway collisions in dense star clusters. Light seeds reach the observed masses only through sustained near-Eddington or super-Eddington accretion \citep{Chaikin2026, Prole}, plausibly favoured in actively merging, shock heated environments \citep{Ambra}, and \citet{Jeon2026} argue from halo abundance alone that cold-mode inflow is required to supply it; heavy seeds relax the timing constraint but demand rare, finely balanced formation sites \citep{Chon}. The question of the dominant channel remains open: the physically relevant prescriptions are not settled \citep{Inayoshi2020, Habouzit, Regan2024, Liu&Inayoshi}, direct constraints at $z \gtrsim 5$ are sparse \citep{Brooks, Roberts}, and the black hole mass function measured over $3.5 < z < 6$, now reaching below $\log_{10}(M_\text{BH}/\text{M}_\odot) = 7$, is a largely featureless power law, from which \citet{Taylor2025} conclude that signatures of seeding have been lost by $z \simeq 5-6$. Whether the seeding history leaves a recoverable signature at all is a question requiring a model in which the seed mass is allowed to depend on environment.

Contemporary simulations are poorly equipped to confront these constraints, given that the black holes themselves are frequently absent. Many of the flagship radiation-hydrodynamical simulations of the reionisation era omit black hole physics entirely for computational economy, including \textsc{Sphinx} \citep{Sphinx}, \textsc{Cosmic Dawn III} \citep{Lewis} and \textsc{Thesan-Zoom} \citep{ThesanZoom, McClymont}. Large volume simulations which do include black holes typically seed a single fixed mass in every halo crossing a threshold, erasing the diversity of formation channels by construction \citep{Habouzit}.

\Acp{sam} are a natural instrument for exploring that diversity, since seeding prescriptions can be exchanged and re-tuned at low computational cost over large volumes. \citet{Jeon}, for example, combine light and heavy seed channels within the \textsc{A-Sloth} model of early star formation to predict the black hole mass function at $z \gtrsim 9$. \citet{Spinoso2023} follow four distinct seeding channels through \textsc{L-Galaxies} on the \textsc{Millennium-II} trees. \citet{Dayal} sets light, heavy and primordial seeds against the same observables and finds that only light seeds growing at the Eddington limit are excluded. \citet{Chen2025} carry the irradiation, enrichment and dynamical heating a halo experiences onto cosmological trees, concluding that these environmental processes act divergently on the masses of the seeds which result. Among hydrodynamical simulations, the \textsc{Brahma} suite \citep{Bhowmick2024} systematically varies gas-based seeding criteria within the \textsc{IllustrisTNG} galaxy formation model.

Current prescriptions nevertheless share a structural limitation: seeding is channel-based, with each channel injecting seeds at a fixed characteristic mass, and so imprinting artificial peaks on the resulting mass function. When environment contributes to those channels, close halo pairs supply the Lyman-Werner flux that a direct collapse requires \citep{Dijkstra2008, Agarwal2012}. The irradiation and enrichment a halo inherits select among light, medium-weight and heavy seeds \citep{Sassano2021, Spinoso2023}. \textsc{Brahma} suppresses seeding in galaxies with fewer than two neighbouring haloes more massive than the host within five virial radii. The suppression recovers the small scale clustering which a galaxy mass criterion alone underestimates \citep{Bhowmick2024}. In each case, environment determines which channel operates, and the mass which follows belongs to the channel rather than to the surroundings. Even where environment is allowed to reach the mass, it does so by delaying collapse until a different stellar evolutionary channel fires \citep{Chen2025}. The seed mass spectrum which results is therefore bimodal, and the grading is discrete. No existing prescription makes the seed mass itself a continuous function of the environment in which the halo forms.

The authors of these prescriptions recognise the limitation: the channel mass ranges are ``not yet firmly assessed and may partly overlap, forming a continuum distribution'' \citep{Sassano2021}. The route offered towards larger volumes, however, is a fit to the direct collapse fraction in redshift alone, proposed as a subgrid prescription for cosmological simulations \citep{Agarwal2012}. The fit retains none of the spatial information from which it was derived. Where neither positions nor gas structure are available, as in the extended Press-Schechter trees used by \citet{Jeon}, the haloes hosting super-Eddington growth cannot be identified at all. Such growth is then assigned at random, as it is by \citet{Bonoli2025}, who draw the rate itself from a power law. The branches they graft below the resolution limit carry a radius from their host but no position of their own.

There are, meanwhile, concrete grounds, physical and now observational, to expect the large scale environment to regulate both seeding and early growth. Heavy seed formation is itself an environmental phenomenon, relying on near-critical irradiation from star forming neighbours or on the dynamical heating of rapidly assembling regions \citep{Regan2024}. Overdense regions assemble structure earliest, host the highest merger rates, and sustain the directed inflows which feed efficient accretion. Megaparsec scale environment measurements around faint \acp{agn} at $z \sim 5$ now furnish independent evidence for overmassive black holes in low mass galaxies, and for a black hole-halo mass relation at these epochs \citep{Matthee2025}. The same comparison at low redshift, against inactive galaxies matched in stellar mass and star formation rate, finds host haloes indistinguishable from the controls at moderate black hole masses and only tentative evidence of more massive haloes at the highest \citep{Mountrichas2026}, which places the discriminating power of such measurements at early epochs. \citet{Chittenden} introduced the radial mass-weighted skewness $\mu_3$, tracing asymmetric, merger-prone mass distributions, and, in \textsc{IllustrisTNG}, showed it to be predictive of chemical enrichment histories from the dark matter component alone. Applied to \textsc{Thesan}, the same quantity singles out the hosts of the most massive early black holes: the quiescent galaxies in the feedback experiments of \citet{Chittenden3} occupy the densest environments in the volume, with persistently elevated radial skew tracing the sustained, anisotropic inflow that fuels their prolonged black hole growth. The same association is found in \textsc{Colibre}, whose \acp{mqg} inhabit more overdense regions on sub-megaparsec and cosmic web scales well before they quench \citep{ChandroGomez2}. Environment is thus measurable from the dark matter alone and, within current simulations, predictive of where the earliest black holes assemble.

In this paper, we introduce \textsc{Aurora}, a black hole seeding model in which the large scale environment, in conjunction with the host halo mass, sets the seed mass directly. Each eligible halo receives a seed determined by a seeding law, log-linear in overdensity, radial skewness and halo mass, producing a continuous, unimodal spectrum of seed masses in this cosmology which replaces the light/heavy dichotomy with a single, smoothly varying population. We ask whether seeding of this kind leaves a measurable signal in which galaxies quench, and in which environments. The \textsc{Aurora} code also contains two optional modifiers: a logistic merger-driven boost \citep{Prole2024, Ambra} and a capped, Eddington-limited growth burst.

\textsc{Aurora} is implemented as a module of \textsc{Mimic}\footnote{\hyperlink{https://github.com/darrencroton/mimic}{https://github.com/darrencroton/mimic}} (Croton et al. in prep), a modular semi-analytic framework, within which the adopted baryonic physics is the \textsc{Sage} \acl{sam} \citep{Croton2006, Croton2016}. Unlike prescriptions keyed to gas metallicity or to the local radiation field, \textsc{Aurora} requires only halo masses and positions, and so needs no baryonic information from the host simulation. Furthermore, given that its environment statistics are precomputed once per snapshot with a periodic $k$-d tree (\Cref{sec:aurora_cache}) rather than evaluated per galaxy, the cost of applying it scales as $\mathscr{O}(N\log N)$ with the size of the halo catalogue. The same physics can therefore be applied across the full dynamic range of a modern simulation suite: from the coarse \textsc{Millennium} data used here \citep{Springel_2005} to simulations of higher mass resolution, where resolution-sensitive objects of interest can be simulated in detail.

This paper is organised as follows. \Cref{sec:mimic-aurora} presents the model: the \textsc{Mimic} framework in which \textsc{Sage} and \textsc{Aurora} are embedded (\cref{sec:mimic}), and \textsc{Aurora} itself (\cref{sec:aurora}), covering its environment metrics, their precomputation and caching, the seed mass spectrum, and the post-seeding growth phase. \Cref{sec:optimisation} calibrates the prescription against the stock \textsc{Sage} black hole population (\cref{sec:calibration}) and examines the sensitivity of the seed mass spectrum and the $z=0$ black hole mass function to each coupling (\cref{sec:sweeps}). \Cref{sec:evolution} follows the consequences forward in time: the structure of the three couplings (\cref{sec:couplings}), the assembly of massive black holes from their seeds (\cref{sec:synthesis}), and the seed-dominated growth pathway (\cref{sec:pathway}). The same section also explores the environmental dependence of quenching (\cref{sec:quiescence}). \Cref{sec:discussion} discusses the implications for inferring seeding histories, the extension of the prescription to further physics, and its application at high redshift. We summarise in \cref{sec:conc}.

Throughout this work, quantities that are free parameters of \textsc{Aurora} are marked with a check accent (e.g. $\check{\alpha}$, $\check{\sigma}_\delta$), distinguishing them at a glance from measured or derived quantities. Their adopted values are collected in \Cref{tab:auroraparams}.

\begin{table}
\centering
\caption{A comprehensive table of parameters used in the \textsc{Aurora} black hole seeding model and their corresponding optimal values. These parametric quantities are referenced throughout this work, and are marked with a check (i.e. $\check{a}$) in the main text, in order to distinguish them as \textsc{Aurora} parameters. For clarity, the checks are not shown in the notation column of this table. The values are those of the optimised configuration derived in this work: the seed law normalisation and coefficients are calibrated against the stock \textsc{Sage} black hole population, and the skew variance floor is calibrated against the Gaussianity and tracer-support criteria of \Cref{sec:aurora_metrics}. The post-seeding Eddington growth budget is set to a negligible value in this configuration, such that the seeds enter \textsc{Sage}'s standard accretion physics essentially unmodified.}
\label{tab:auroraparams}
\begin{tabular}{@{}lccc@{}}
\hline
Parameter & Notation & Value & Unit\\
\hline
Minimum Halo Particles & $N_h^\text{min}$ & 115 & ---\\
Satellite Mass Ratio Minimum & $q_\text{sat}$ & 0.1 & ---\\
Satellite Velocity Maximum & $v_\text{sat}$ & 1.5 & $V_\text{vir}$\\
Specific Seed Normalisation & $M_\bullet^0$ & $3.587\times10^{-6}$ & ---\\
Overdensity $(\delta)$ Coefficient & $\alpha$ & 1.3 & ---\\
Skew $(\mu_3)$ Coefficient & $\beta$ & 0.4 & ---\\
Halo Mass $(M_h)$ Coefficient & $\gamma$ & 0.275 & ---\\
Merger Boost Maximum & $B_\text{max}$ & 2 & ---\\
Merger Boost Steepness & $k$ & 1 & ---\\
Merger Boost Skew Threshold & $\mu_3^\text{th}$ & 2 & ---\\
Number of Salpeter Times & $N_\text{Salp}$ & $10^{-6}$ & ---\\
Eddington Radiative Efficiency & $\varepsilon$ & 0.1 & ---\\
Minimum Seed Fraction & $f_\text{min}$ & $2\times10^{-8}$ & ---\\
Maximum Seed Fraction & $f_\text{max}$ & $2\times10^{-3}$ & ---\\
Smoothing Scale (Overdensity) & $\sigma_{\delta}$ & 1 & cMpc/$h$\\
Smoothing Scale (Skew) & $\sigma_{\mu_3}$ & 1 & cMpc/$h$\\
Gaussian Kernel Cutoff & $N_\text{cut}$ & 3 & ---\\
Skew Variance Floor & $F_{\mu_3}$ & 0.2 & (cMpc/$h$)$^2$\\
Skew Variance Floor Index & $p_{\mu_3}$ & 2 & ---\\
\hline
\end{tabular}
\end{table}

\section{Model Setup}
\label{sec:mimic-aurora}

All results in this work derive from the \textsc{Millennium} simulation \citep{Springel_2005}: a periodic, pure dark matter volume of $1.25 \times 10^8\, (\text{cMpc}/h)^3$, with 64 snapshots from $z=127$ to $z=0$. Its key properties are collected in \cref{tab:minimill}. Haloes are identified with the \ac{fof} and \textsc{Subfind} \citep{Springel_2001} algorithms and linked into \textsc{L-Halo} merger trees, which \textsc{Mimic} reads natively.

\begin{table}
\centering
\caption{Properties of the \textsc{Millennium} simulation \citep{Springel_2005} used
throughout this work, including aspects of the simulation's size, resolution and cosmology.}
\label{tab:minimill}
\begin{tabular}{lccc}
\hline
Property & Notation & Value & Unit\\
\hline
Box side length & $L_\text{box}$ & $500$ & $\text{cMpc}/h$\\
Particle number & $N_\text{part}$ & $2160^3$ & ---\\
Particle mass & $m_p$ & $8.6\times10^{8}$ & $\text{M}_\odot/h$\\
Force softening & $\epsilon$ & $5$ & $\text{ckpc}/h$\\
Snapshots & $N_\text{snap}$ & 64 & ---\\
Matter density & $\Omega_\text{m}$ & 0.25 & ---\\
Baryon density & $\Omega_\text{b}$ & 0.045 & ---\\
Dark energy density & $\Omega_\Lambda$ & 0.75 & ---\\
Hubble parameter & $h$ & 0.73 & ---\\
Fluctuation amplitude & $\sigma_8$ & 0.9 & ---\\
Spectral index & $n_s$ & 1 & ---\\
\hline
\end{tabular}
\end{table}

\subsection{The \textsc{Mimic} Semi-Analytic Framework}
\label{sec:mimic}

\textsc{Aurora} is implemented as a module within \textsc{Mimic}: a semi-analytic galaxy evolution framework built around a physics-agnostic core (Croton et al. in prep). The defining architectural choice is a strict separation of three concerns which traditional \acp{sam} compile together. The framework core handles merger tree processing, memory management, input/output, and output provenance, and has no compiled-in knowledge of any specific physics. The astrophysics is instead defined within self-contained \emph{model packages}: collections of runtime modules such as star formation, supernova feedback, \ac{agn} heating, black hole growth, and \textsc{Aurora} itself, which register with the framework at startup and communicate exclusively through a shared galaxy property system. The simulation input is modular in the same way: \emph{simulation packages} wrap each merger tree catalogue with its cosmology, units, and on-disk field layout. Changing the physics, the module ordering, or the parameter values, requires editing a YAML configuration file rather than recompiling, such that the same physics runs unchanged across different simulations.

The fiducial baryonic physics used in this work is the \textsc{Sage} model \citep{Croton2006, Croton2016}, provided as \textsc{Mimic}'s \texttt{sage16} package, a modular reproduction of the published code. \textsc{Aurora} is deliberately kept independent of this; as a \textsc{Mimic} module, \textsc{Aurora} depends only on halo structural quantities, namely halo mass, position, and halo type, together with the shared black hole mass property. \textsc{Aurora} declares no dependency on any \textsc{Sage} module or parameter, and writes its output to its own dedicated seeding properties alongside that single shared channel.

The runs presented here combine the two as a single configuration, registered for convenience as the \texttt{sage16-aurora} package. This is a runtime choice rather than a structural coupling: \textsc{Aurora} seeds the black hole mass, and whichever growth model is active, \textsc{Sage} in this case, subsequently evolves the black holes from \textsc{Aurora}'s seeds. Disabling \textsc{Aurora} recovers the original \textsc{Sage} model exactly, and \textsc{Aurora} itself could be combined with a different baryonic model with minimal change, since its only shared surface is \textsc{Mimic}'s property system. This independence is mutual and deliberate, allowing the seeding prescription to be developed and tested without entanglement in the surrounding galaxy formation physics, and allowing the host physics to run with or without it.

In the practice of optimising parameters and physical prescriptions to observational constraints with most traditional models, one would typically devise a framework requiring recompilation for each new seeding mode, or embedded cross-module dependencies in shared headers, imposing a prohibitive iteration cost. In \textsc{Mimic}, adding a new seeding mechanism requires only a module source file and a metadata description; every other component (the tree traversal, the \textsc{Sage} baryonic physics, the output system and the \textsc{Aurora} environment cache) continues to operate without modification.

\subsection{\textsc{Aurora}: Environment-Dependent Black Hole Seeding}
\label{sec:aurora}

\textsc{Aurora} is a black hole seeding module which assigns seed masses to dark matter haloes based on their local large scale environment, characterised by two Gaussian-smoothed scalar metrics: the matter overdensity $\delta$, and the radial mass-weighted skewness $\mu_3$ introduced by \citet{Chittenden}. The module runs as a pre-timestep, full-halo module: it executes once per snapshot interval and receives each complete \ac{fof} group before the baryonic physics is integrated over the substeps of that interval (ten in the runs presented here). This placement guarantees that seeding is complete before any substep touches the black hole mass channel. In the original \textsc{Sage} model, a black hole has no explicit seed: it comes into being with the first quasar-mode accretion event onto a galaxy whose black hole mass is still zero. When \textsc{Aurora} is active, quasar-mode accretion is applied only to galaxies whose black hole mass is already positive, such that every black hole begins as an \textsc{Aurora} seed, and grows thereafter under the unchanged \textsc{Sage} physics.

The environment metrics themselves are not measured during this pass; they are precomputed against the complete halo distribution of each snapshot and looked up here (\Cref{sec:aurora_cache}), ensuring that a halo's seed never depends on which fraction of the volume the depth-first traversal happens to have visited. A halo receives a seed black hole if it hosts a galaxy which does not yet carry a black hole ($M_\text{BH} = 0$), its particle count meets a minimum resolution threshold $\check{N}_h^\text{min}$, and it passes the massive neighbour gate described below. The criterion is structural rather than morphological, and no galaxy type is excluded: a satellite or orphan galaxy which has yet to acquire a black hole is seeded for the same reason as a central.

\textsc{Aurora} does, however, correct for one configuration in which its environment metrics cease to measure the large scale environment. For a halo lying within another's virial radius, the Gaussian kernel of \Cref{sec:aurora_metrics} is dominated by that single host, which sits well inside the smoothing scale, so that $\delta$ and $\mu_3$ record the mass of the group the halo has fallen into rather than the structure around it. Read at face value, the prescription would then plant its largest seeds in the satellites of the most massive groups, whose own supply of gas is being removed by tides and ram pressure stripping. An optional massive neighbour gate, active in the configuration adopted here, therefore withholds a seed from such a halo only where any halo within the aperture dominates in one of two ways: in mass, at $M_\text{vir}/M_\text{vir}^\text{nbr} < \check{q}_\text{sat}$, with $M_\text{vir}^\text{nbr}$ the virial mass of the neighbour, or in dynamics, where the halo moves through the neighbour as a satellite, faster than $\check{v}_\text{sat}$ times the dominant halo's virial velocity, following a plunging orbit on which ram pressure removes that supply. The criterion concerns what the metrics can measure for such a halo, rather than any property its galaxy could sense. Both criteria read only quantities the merger trees already carry, preserving \textsc{Aurora}'s independence from the baryonic model, and both are re-evaluated at every snapshot, so that a backsplash halo which passes through a larger halo becomes eligible once more.

The gate is selective rather than a blanket exclusion of satellites: a halo carrying at least a fraction $\check{q}_\text{sat} = 0.1$ of its massive neighbour's mass, and moving more slowly than $\check{v}_\text{sat} = 1.5$ times the neighbour's virial velocity, remains eligible. Such systems account for the $11$ per cent of the 4.6 million seeding events in this volume which occur in satellites; the remaining $89$ per cent occur in centrals, while no orphan in \textsc{Millennium} happens to meet $\check{N}_h^\text{min}$.

\subsubsection{Environment metrics}
\label{sec:aurora_metrics}

\textsc{Aurora} characterises the local environment of each halo using two complementary statistics, each evaluated over a Gaussian-smoothed neighbourhood.

\paragraph*{Overdensity:}
The Gaussian-smoothed local mass density of a halo is
\begin{equation}
  \rho = \frac{\sum_j M_j\, \Upsilon(r_j, \check{\sigma}_\delta)}
              {\left(2\pi\check{\sigma}_\delta^2\right)^{3/2}},
\end{equation}
where the sum excludes the target halo itself, $r_j$ is the periodic minimum-image distance to neighbour $j$, $M_j$ is its halo mass, and $\Upsilon(r,\sigma) = \exp(-r^2/2\sigma^2)$ is a Gaussian weight of unit peak amplitude, truncated at $r = \check{N}_\text{cut} \times \check{\sigma}_\delta$, so that $\check{\sigma}_\delta$ sets the smoothing scale and $\check{N}_\text{cut}\check{\sigma}_\delta$ defines the aperture within which neighbours contribute. The weight alone does not integrate to unity over the volume; the $(2\pi\check{\sigma}_\delta^2)^{3/2}$ prefactor supplies that normalisation, making $\rho$ a mass density rather than a weighted-mean mass.

The fractional overdensity is then
\begin{equation}
  \delta = \frac{\rho - \bar{\rho}}{\bar{\rho}} = \frac{\rho}{\bar{\rho}} - 1,
\end{equation}
where, in contrast to a cosmic-mean reference, $\bar{\rho}$ is the \emph{geometric} mean of $\rho$ taken over all haloes at the same snapshot. This population-relative reference is computed once per snapshot during cache generation and stored in the cache header. Using the geometric mean ensures that the logarithmic overdensities are zero-centred across the halo population, and it avoids the vanishing denominators that a volume-averaged matter density produces at sparsely resolved high redshift epochs. Haloes with fewer than five contributing neighbours within the kernel are assigned a neutral $\delta = 0$ and excluded from the geometric mean reference.

\paragraph*{Radial Skewness:}
The radial skewness $\mu_3$ is the third mass-weighted standardised moment of the neighbour distance distribution, introduced in \citet{Chittenden} as a measure of the dominant interactions of an environment with the target halo. Using weights $w_j = M_j\,\Upsilon(r_j, \check{\sigma}_{\mu_3})$, the mass-weighted moments are defined as follows:
\begin{align}
  \mu_1 &= \frac{\sum_j w_j\, r_j}{\sum_j w_j},
  \label{eq:skew}\\
  \mu_2 &= \frac{\sum_j w_j\, (r_j - \mu_1)^2}{\sum_j w_j},
  \nonumber \\
  \mu_3 &= \frac{\sum_j w_j \left(\dfrac{r_j - \mu_1}{\sqrt{\mu_2}}\right)^3}
                {\sum_j w_j}.
  \nonumber
\end{align}
A positive skewness indicates neighbouring mass concentrated close to the target halo, with a long tail of distant neighbours, signifying an asymmetric mass distribution associated with merger-prone and filamentary environments. A negative skewness indicates mass concentrated far from the halo, with a tail of near neighbours \citep[see fig. 1 of][]{Chittenden}. Haloes with fewer than five contributing neighbours within the kernel are assigned $\mu_3 = 0$.

Standardised moments of this kind possess a structural failure mode: where the kernel weight is carried by only one or two contributors, $\mu_2$ collapses and $\mu_3$ approaches a near-singularity, reaching values of order $\pm 10$ whose sign and magnitude are noise-dominated. Raw neighbour counts do not detect this; the median offending evaluation in \textsc{Millennium} has sixteen in-kernel neighbours, of which the two heaviest carry 88 per cent of the weight. \textsc{Aurora} therefore regularises the estimator with an adaptive variance floor,
\begin{equation}
  \label{eq:skewfloor}
  \mu_2 \rightarrow \mu_2 + \frac{\check{F}_{\mu_3}}{N_\text{eff}^{\,\check{p}_{\mu_3}}},
  \qquad
  N_\text{eff} = \frac{\left(\sum_j w_j\right)^2}{\sum_j w_j^2},
\end{equation}
where $N_\text{eff}$ is the effective, weight-based contributor count of the kernel. The floor is negligible for well-supported evaluations ($N_\text{eff} \gg 1$) and reaches full strength as $N_\text{eff} \rightarrow 1$. We adopt $\check{F}_{\mu_3} = 0.2\,(\text{cMpc}/h)^2$ and $\check{p}_{\mu_3} = 2$, calibrated to bring the $\mu_3$ population as close to Gaussian as possible while retaining extreme values only where the kernel is well traced: at this optimum the Anderson-Darling non-Gaussianity statistic falls by nearly two orders of magnitude, while well-supported values shift by less than five per cent. Setting $\check{F}_{\mu_3} = 0$ recovers the unregularised estimator.

The regularisation reflects how \textsc{Aurora} uses the statistic rather than any defect in its original formulation. In \citet{Chittenden}, $\mu_3$ entered a semi-recurrent neural network as a predictive feature constraining chemical enrichment histories, in an approach carried unchanged to pure dark matter simulations by \citet{Chittenden2} and relying on the distribution and numerical accuracy of skew values across all mass regimes and cosmic epochs. These works evaluated $\mu_3$ without a smoothing kernel and over the denser catalogues of higher resolution suites, where the failure mode is scarce. Replacing the Gaussian kernel with a top-hat aperture halves the incidence of $|\mu_3| > 3$, whereas diluting the \textsc{Millennium} catalogue fourfold, a crude inversion of the move to higher resolution, drives the rank-deficient ($N_\text{eff} < 2$) fraction from $9$ to $44$ per cent. More importantly, the predictions there were decided by the historical variation of the skew in conjunction with the halo's evolutionary history, leaving an aberrant evaluation at any single epoch buffered by the rest of the sequence. \textsc{Aurora} instead reads one instantaneous value at the moment of seeding, through a mass-weighted Gaussian kernel which concentrates weight on the nearest massive neighbours, over the coarser catalogues typical of large volume dark-matter-only runs, and exponentiates it into a seed mass. Thus, \Cref{eq:skewfloor} restores the statistic's behaviour without altering its definition.

\begin{figure}
\includegraphics[width=\columnwidth]{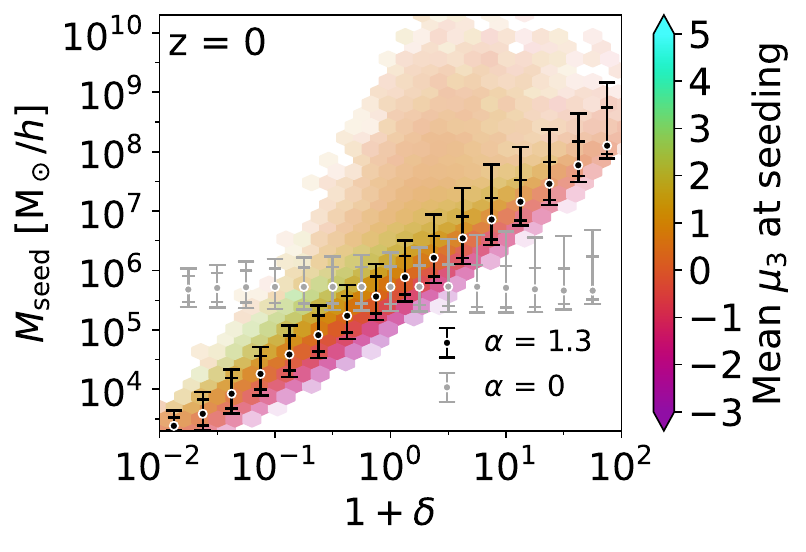}
\caption{A 2D hexbin plot showing the spectrum of \textsc{Aurora}-generated black hole seed masses (vertical axis) as a function of overdensity (horizontal axis) and skew (colour scale), both measured with a Gaussian kernel of $1\,\text{cMpc}/h$ truncated at $3\,\text{cMpc}/h$, in the optimised configuration of \cref{tab:auroraparams}. The opacity of each hex bin indicates the relative occupancy of said bin, while the black points and error bars show the median, $5$th-$95$th percentile range, and $1$st-$99$th percentile range of seed masses, in logarithmically spaced overdensity bins, after removing the local slope of the median relation within each bin. The grey points and error bars correspond to the run where the dependence on overdensity is removed, i.e. $\check{\alpha}=0$, in bins offset from the optimised run ($\check{\alpha}=1.3$) by half a bin width, such that the two sets alternate. The diffuse "cloud" above the main relation holds just 0.6 per cent of the seeds shown, namely those lying more than 0.65 dex above the median at their overdensity. Four fifths of them were planted in haloes growing rapidly enough to exceed the threshold mass between snapshots, with extreme skew values supplying the rest.}
\label{fig:seedmassspec}
\end{figure}

The two smoothing scales, $\check{\sigma}_\delta$ and $\check{\sigma}_{\mu_3}$, are independent free parameters. They need not be equal: a larger $\check{\sigma}_\delta$ captures large scale density waves, while a smaller $\check{\sigma}_{\mu_3}$ is more sensitive to nearby merging companions. In this work both are set to $1\,\text{cMpc}/h$. The kernel is truncated at $\check{N}_\text{cut}$ times the relevant smoothing scale, giving an aperture of $3\,\text{cMpc}/h$ for both metrics\footnote{Evaluating $\delta$ and $\mu_3$ with apertures from $1-5\,\text{cMpc}/h$ yielded similar distributions, and so only one aperture was chosen for this work, as the same metrics with additional apertures were effectively redundant as seed mass predictors.} in the configuration of \cref{tab:auroraparams}, as quoted alongside the environment metrics in \cref{fig:seedmassspec}. All distance calculations use the minimum-image convention in a periodic box.

\Cref{fig:seedmassspec} shows the seed mass spectrum which results from the configuration optimised to the stock \textsc{Sage16} model (see \cref{sec:optimisation}), as a joint function of the two environmental metrics: seed mass increases with overdensity across some four decades in $1+\delta$, with the radial skewness modulating the relation at fixed density. Without the overdensity coupling, the median seed mass varies by less than $0.1$ dex across the full range of $1+\delta$, showing that the trend is not inherited from the host haloes. \Cref{fig:seedmassspec} does not correspond to a snapshot; it pools every seeding event in the volume across all $64$ snapshots, with each halo entering exactly once, at the moment of seeding. The quantity plotted is the mass assigned at that instant, from which the black hole subsequently grows under \textsc{Sage}'s accretion physics. The continuous, unimodal spectrum this produces replaces the discrete channel masses of conventional prescriptions. The remainder of this section will illustrate how this relation is established.

\subsubsection{Environment precomputation and caching}
\label{sec:aurora_cache}

Evaluating the environment metrics requires the positions and masses of every halo present at a snapshot, not only those of the \ac{fof} group currently being processed. \textsc{Mimic} can traverse the merger trees vertically, evolving each tree in full to the present before reading the next, as \textsc{Sage} does, or horizontally, advancing every tree through one snapshot at a time. Under vertical traversal, the stream returns to early snapshots with each new tree, so that a halo sees only those others belonging to trees already traversed, and a metric accumulated on the fly would be evaluated against an incomplete and growing subset of the population. Under horizontal traversal, the snapshot population is complete, but the metrics are still required before the group carrying them is advanced. \textsc{Aurora} therefore precomputes the environment in a dedicated pass in either case: it ingests every halo of every tree into per-snapshot catalogues of masses and positions, evaluates the metrics for each halo against the complete catalogue for its snapshot, and writes one cache file per snapshot. The main run reads these values directly, and so is correct on its first and only pass, loading one snapshot's cache at a time and querying it in constant time.

Separating the expensive neighbour computation from the seeding itself is central to \textsc{Aurora}'s scalability and memory footprint. The basic prescan is a brute force $\mathscr{O}(N_\text{cat}^2)$ evaluation which holds every per-snapshot catalogue in memory at once, and so is comfortable for test volumes but infeasible for larger simulation datasets. For these, the caches are instead generated ahead of the run by a standalone prepass that builds the same statistics snapshot by snapshot with a periodic $k$-d tree, reducing the cost to $\mathscr{O}(N\log N)$ and parallelising across snapshots. Because the environment metrics are defined per snapshot, and are independent of merger history, this precomputation is agnostic to which of the two traversal modes the consuming run uses: the prepass writes byte-identical cache files regardless, and a run then consumes them directly while the in-engine prescan detects their presence, bypassing the calculation if the cache files already exist. This allows identical seeding physics to run at a fraction of the time and memory of the full computation.

\subsubsection{Seed mass functions}
\label{sec:aurora_seedmass}

\textsc{Aurora} produces a continuous seed mass spectrum, rather than the binary choice between light and heavy seeds of fixed masses adopted in most contemporary models \citep[e.g.][]{Jeon}. While several simplistic seed mass functions were tested during \textsc{Aurora}'s development, and remain implemented in the \textsc{Aurora} code, the scientific mode of operation produces a unimodal, continuously varying seed mass spectrum:
\begin{align}
\label{eq:BHseeding}
\ln \left( \frac{M_\bullet}{N_p\, m_p} \right) = & \ln \left( \check{M}_\bullet^0 \right) \\
+ & \ \check{\alpha} \times \ln \left( 1 + \delta \right) \nonumber \\
+ & \ \check{\beta} \times \mu_3 \nonumber \\
+ & \ \check{\gamma} \times \ln \left( \frac{M_h}{\check{N}_h^\text{min}\, m_p} \right) \nonumber
\end{align}
The law is formulated in terms of the specific seed mass: the seed expressed as a fraction of the mass which made the halo eligible, $N_p\, m_p$, with $N_p$ the halo's particle count and $m_p$ the simulation particle mass\footnote{This form assumes a single particle mass, as in the uniform-resolution simulations considered here. In a multi-mass or zoom-in simulation the eligibility mass would instead be the summed particle mass of the halo; the law is otherwise unchanged.}. $\delta$ is the normalised overdensity of \Cref{sec:aurora_metrics}, and $\check{M}_\bullet^0$ is the dimensionless specific normalisation: the seed-to-eligibility-mass ratio assigned to a threshold mass halo in a mean density ($\delta = 0$), symmetric ($\mu_3 = 0$) environment. The halo mass term, which compares the virial mass $M_h$ with the threshold mass $\check{N}_h^\text{min}\, m_p$, is likewise small for a halo which crosses the threshold with little overshoot, since the two then lie within a few tens of per cent of each other, so that $\check{\gamma}$ acts appreciably only on haloes which overshoot it between snapshots (\Cref{sec:couplings}). The result is clamped between $\check{f}_\text{min} \times N_p\, m_p$ and $\check{f}_\text{max} \times N_p\, m_p$. Both bounds are relative, scaling with the eligibility mass of whatever halo crosses the threshold: no halo can receive a seed exceeding the fraction $\check{f}_\text{max}$ of the mass that made it eligible, but the ceiling is not an absolute one. At the resolution threshold it stands at $\check{f}_\text{max}\check{N}_h^\text{min} m_p \simeq 2\times10^{8}\,\text{M}_\odot/h$, and it rises in proportion for any more massive crosser, so the prescription does not of itself exclude seeds heavier than any physical formation channel produces. Parametric values are listed in \cref{tab:auroraparams}.

The three environment and mass terms contribute independently and additively in logarithmic space, which has the property that each is approximately normally distributed: overdensity is approximately lognormal, skew is approximately normal, and halo mass is distributed by the halo mass function by definition. The seed mass distribution is thus approximately lognormal, albeit with a positive tail determined by the highest values of the environmental metrics.

Two features distinguish \Cref{eq:BHseeding} from channel-based prescriptions: the continuity of the spectrum it produces, and the coupling of that spectrum to environment. Their effects are separable. The scrambled-environment control of \Cref{sec:evolution} retains the seed mass spectrum, together with its dependence on halo mass and epoch, but destroys the correspondence between each seed and the environment in which it formed, so that any signal which survives the control owes to the spectrum alone, and any which does not owes to the environmental coupling.

Following \citet{Prole2024}, the seed mass is further subject to a merger-driven boost, capturing the enhanced early growth expected in actively merging, shock heated environments \citep{Ambra}. The boost is a smooth logistic function of the radial skew,
\begin{equation}
  \label{eq:mergerboost}
  B(\mu_3) = 1 + \frac{\check{B}_\text{max} - 1}{1 + \exp\!\left[ -\check{k}\left(\mu_3 - \check{\mu}_3^\text{th}\right) \right]},
\end{equation}
applied multiplicatively to give the final seed mass $M_\bullet \times B(\mu_3)$. It rises from unity well below the skew threshold $\check{\mu}_3^\text{th}$ to an asymptotic maximum $\check{B}_\text{max}$ for strongly merger-prone haloes, with $\check{k}$ setting the transition sharpness. Because this is a smooth, monotonic function of $\mu_3$, it preserves the continuity and unimodality of the spectrum. The boost is active in the configuration adopted here, with $\check{B}_\text{max} = 2$ following \citet{Prole2024}, so that a seed forming in a strongly merger-prone environment is twice that of an identical halo in a symmetric one. Of all seeding events, 2.3 per cent lie above the threshold $\check{\mu}_3^\text{th}$, and the mean boost across the seeded population is a factor of $1.21$.

With a non-zero direct skew coupling, the radial skewness therefore reaches the seed mass by two routes of very different character: exponentially through \Cref{eq:BHseeding}, and in bounded, threshold-like form through the boost. The first of these is what makes the regularisation of \Cref{eq:skewfloor} consequential rather than cosmetic. At $\check{\beta} = 0.4$ an unregularised excursion to $|\mu_3| \sim 10$ would scale a seed by a factor of $e^{4} \approx 55$, and would simultaneously drive the logistic hard against its asymptote, on the strength of one or two neighbours. The floor ensures that both routes respond to genuine environmental asymmetry.

\subsubsection{Post-seeding Eddington growth}
\label{sec:aurora_eddington}

\textsc{Aurora} optionally follows seed assignment with a short phase of Eddington-limited exponential growth, integrated at a thin-disc radiative efficiency $\check{\varepsilon}$ up to a cumulative budget of $\check{N}_\text{Salp}$ Salpeter times \citep{Salpeter}. The budget is cumulative rather than imposed per step, which makes the total growth independent of the snapshot cadence of the underlying $N$-body simulation; the accreted gas is drawn from the cold reservoir, and growth is gas-limited where that reservoir is insufficient. Because the burst runs in the pre-timestep phase, any cooling, star formation or accretion modules which run subsequently encounter the post-burst black hole mass from the start of the substep loop, with no double counting. In the configuration used here, these are the corresponding \textsc{Sage16} modules, but the ordering guarantee is a property of the framework rather than of that choice. The budget adopted here is negligible (\cref{tab:auroraparams}), so the results presented in this work isolate the seeding prescription itself; we defer the calibration of this growth channel to future work.

\section{Parameter Optimisation and Sensitivity}
\label{sec:optimisation}

\subsection{Calibration against the base model}
\label{sec:calibration}

\textsc{Aurora} introduces environmental structure into black hole seeding; it should do so without degrading the calibrated galaxy population of the base model. We therefore anchor its parameters by requiring that the $z=0$ black hole population of the stock \textsc{Sage} model is recovered when \textsc{Aurora} replaces the stock seeding channel entirely. Because black hole growth is gated on the presence of a seed (\Cref{sec:aurora}), occupation is fixed by the eligibility threshold together with the massive neighbour gate of \Cref{sec:aurora}: the set of $z=0$ black hole hosts is the set of surviving lineages whose haloes ever reach $\check{N}_h^\text{min}$ particles without the gate withholding their seed. The threshold is set first, by count matching: $\check{N}_h^\text{min} = 115$ yields 97 per cent of the black hole hosts in \textsc{Sage}, with the residual attributable to lineages seeded by the stock channel below the resolution floor.

The remaining parameters shape the seed mass spectrum. Their optimisation exploits the gate a second time: once a black hole exists, the growth physics is identical to the base model, so the $z=0$ mass of a threshold crosser in a gated run is predicted, to leading order, by the base run itself,
\begin{equation}
  \label{eq:predictor}
  M_\text{BH}^\text{pred}(z{=}0) \simeq \max\!\left(M_\text{b} - M_\text{e},\, 0\right) + M_\bullet,
\end{equation}
where $M_\text{b}$ is the lineage's $z=0$ black hole mass in the base run, $M_\text{e}$ is its base-run mass one snapshot before first crossing the threshold, and $M_\bullet$ is the \textsc{Aurora} seed evaluated from the environment at the crossing. Evaluating a candidate parameter set thus reduces to a vectorised histogram over the crossing table rather than a full model run, yielding a saving of some six orders of magnitude in cost, which permits an exhaustive grid search.

The objective is a Poisson-style $\chi^2$ between the predicted and stock $z=0$ black hole mass functions in $0.4$ dex bins, augmented by a penalty on the overmassive tail: the number of objects with $M_\text{BH}/M_h > 10^{-3}$ may not exceed the stock model's own count. The grid spans the normalisation and the three coupling indices, with two alternative ceiling conventions: proportional to the halo mass at seeding, or to the eligibility mass $N_p\,m_p$, which converge to the same optimum. The latter, adopted here, is the specific form of \Cref{eq:BHseeding}.

With the adopted bounds (\cref{tab:auroraparams}), applied to the specific seed fraction before the merger boost, the ceiling is inactive in the calibrated configuration, where the largest specific fraction realised is $5.0\times10^{-4}$, a factor of four below $\check{f}_\text{max}$. The floor is mildly active, applying to $1.8$ per cent of seeding events in the calibrated configuration: because $\ln(1+\delta)$ is zero-centred across the halo population by construction (\Cref{sec:aurora_metrics}), the overdensity term lowers the seed mass wherever the environment is underdense, and drives it to the floor in the sparsest. The merger boost is held at unity during this fit, so that the calibration constrains the environmental law in isolation. The resulting values are those of \cref{tab:auroraparams}; confirmation runs of the full pipeline at the optimum reproduce the predicted $\chi^2$ to within $10-20$ per cent.

The three couplings are not constrained equally. The overdensity and halo mass indices are pinned tightly, yet the fit is only weakly sensitive to the skew coefficient over the range $0 \lesssim \check{\beta} \lesssim 0.4$. Since the skew term shifts the mean seed mass as well as reshaping its tail, it trades against the normalisation, and once the normalisation is refit at each value, the objective varies across that range by less than the fidelity of the predictor itself. Larger values are disfavoured, degrading the fit substantially. The adopted $\check{\beta} = 0.4$ therefore sits at the upper end of the permitted plateau, chosen because it produces a visible response in the high mass tail (\Cref{sec:sweeps}) at no significant cost to the $z=0$ black hole mass function. The skew acts on the seed mass both directly through this term and, in bounded form, through the merger boost, as shown in \Cref{eq:mergerboost}.

\subsection{Sensitivity of the black hole population}
\label{sec:sweeps}

\begin{figure}
\includegraphics[width=\columnwidth]{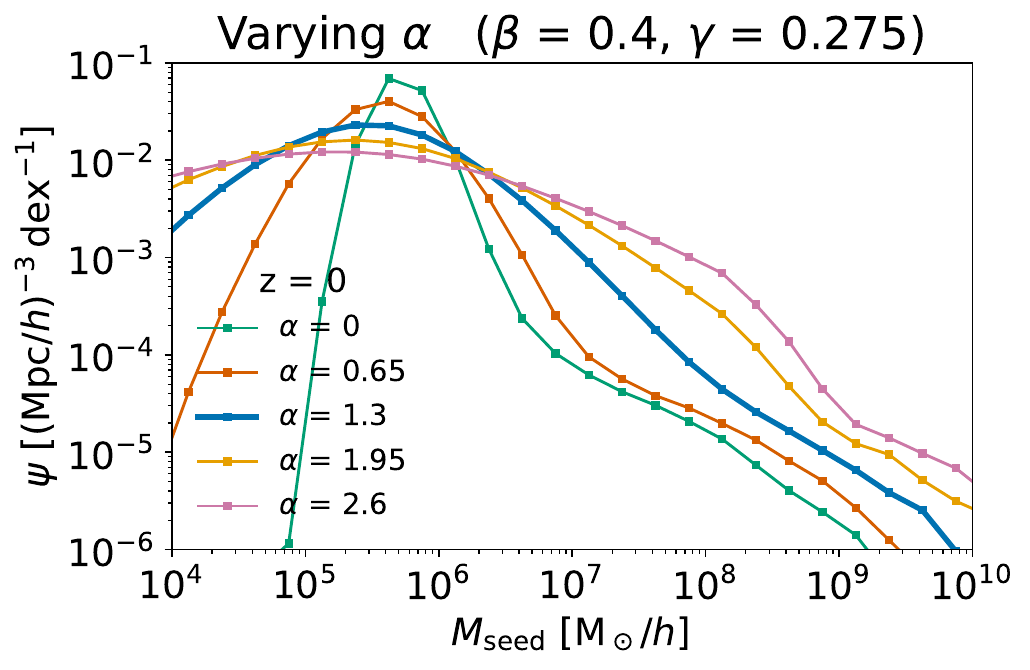}\\
\includegraphics[width=\columnwidth]{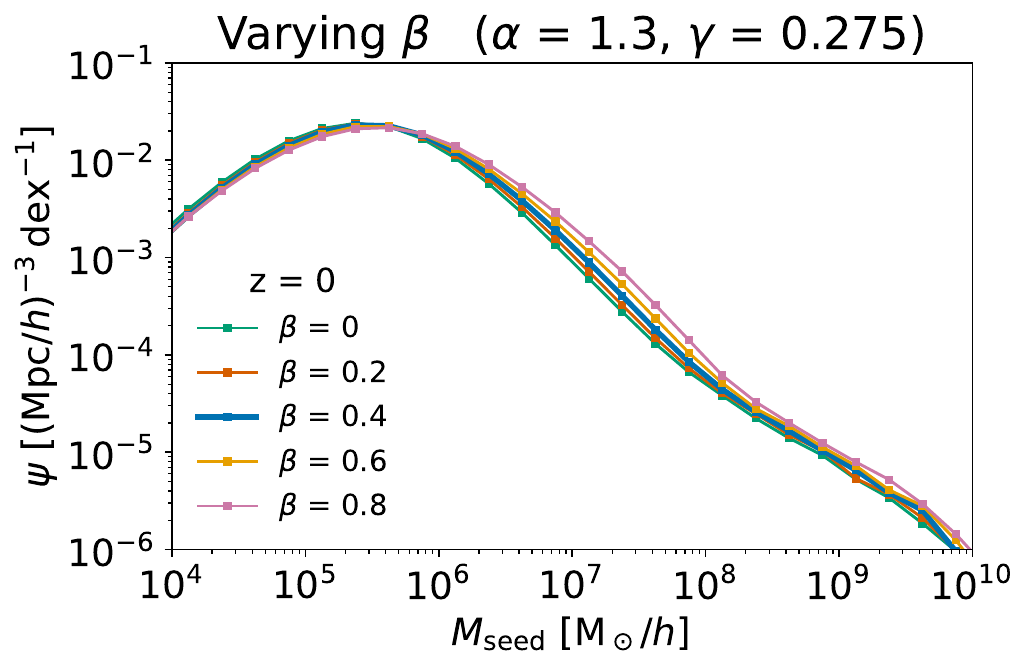}\\
\includegraphics[width=\columnwidth]{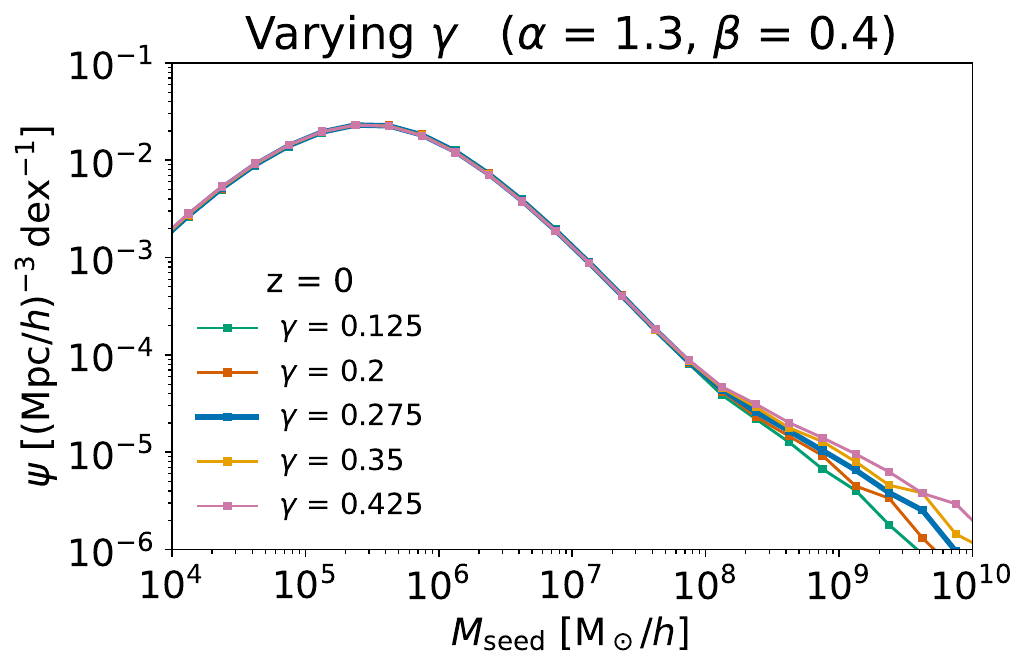}
\caption{Sensitivity of the \textsc{Aurora} seed mass spectrum $\psi$ to the three coupling indices, varied one at a time about the anchor configuration. Top: varying $\check{\alpha}$ (fixed $\check{\beta} = 0.4$, $\check{\gamma} = 0.275$); the spectrum narrows for weak coupling and broadens for strong coupling. Middle: varying $\check{\beta}$ (fixed $\check{\alpha} = 1.3$, $\check{\gamma} = 0.275$); the high mass tail is elevated with little change to the overall shape. Bottom: varying $\check{\gamma}$ (fixed $\check{\alpha} = 1.3$, $\check{\beta} = 0.4$); the bulk of the spectrum is unchanged while the high mass tail separates monotonically, with the coupling acting appreciably only on haloes which cross the eligibility threshold with substantial overshoot. In every panel the optimised configuration of \cref{tab:auroraparams} is drawn in bold. Unlike the black hole mass function $\phi$ (\cref{fig:sweep_bhmf}), the seed mass spectrum is not an observed quantity, and hence there are no survey data to calibrate against.}
\label{fig:sweep_seedspec}
\end{figure}

\begin{figure}
\includegraphics[width=\columnwidth]{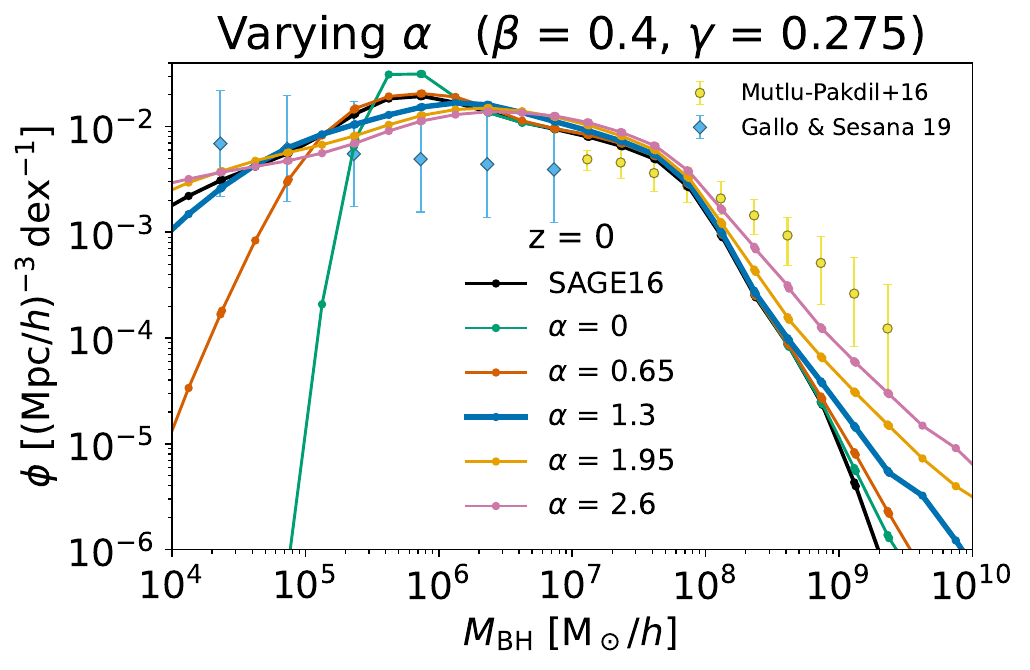}\\
\includegraphics[width=\columnwidth]{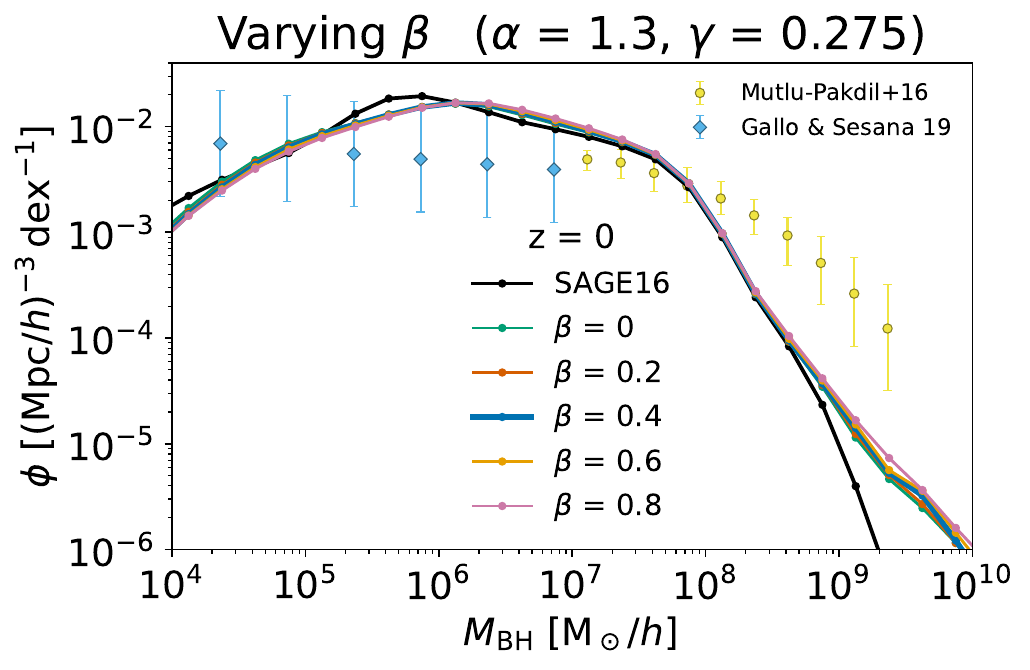}\\
\includegraphics[width=\columnwidth]{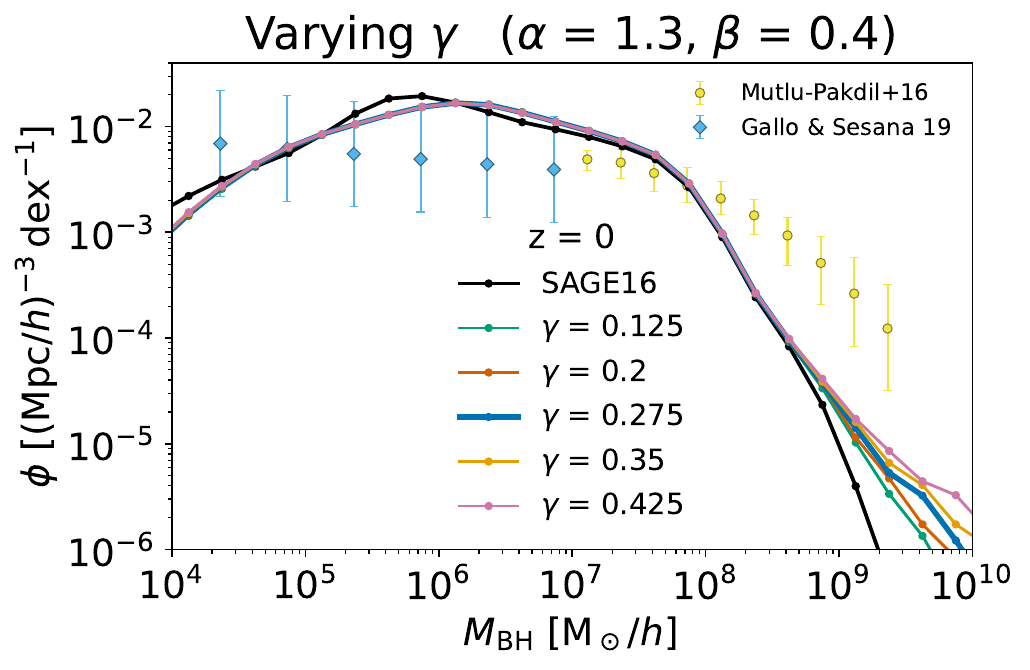}
\caption{The $z=0$ black hole mass function $\phi$ under the same single-parameter variations as \cref{fig:sweep_seedspec} (top to bottom: $\check{\alpha}$, $\check{\beta}$, $\check{\gamma}$), with the stock \textsc{Sage} population in black and the optimised configuration of \cref{tab:auroraparams} in bold in every panel. The divergence at the lowest masses is structural rather than parametric: the stock channel bootstraps black holes in haloes below the eligibility floor, a population the gated model excludes by construction. Published $z=0$ determinations are overlaid: the all-galaxy mass function of \citet{MutluPakdil2016}, tabulated with asymmetric errors, in yellow, and the occupation-corrected estimate of \citet{GalloSesana2019} in sky blue, whose fit we sample at half-decade masses and show with the $68$ per cent normalisation uncertainty those authors quote. We plot the former only above the peak of their determination at a black hole mass of $10^{7.25}\,\text{M}_\odot$.}
\label{fig:sweep_bhmf}
\end{figure}

\begin{figure}
\includegraphics[width=\columnwidth]{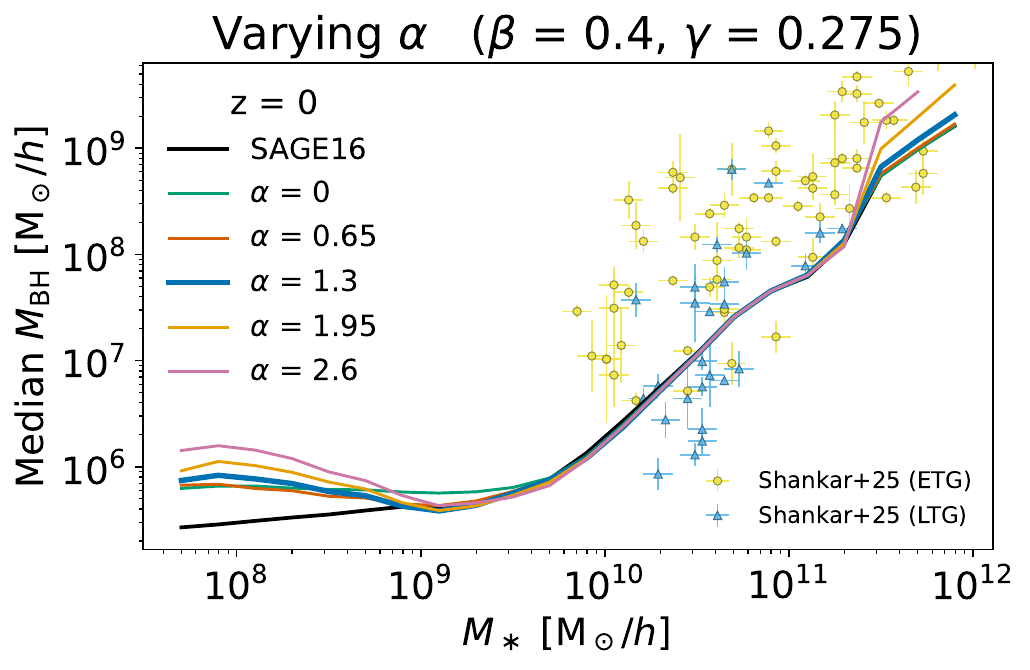}\\
\includegraphics[width=\columnwidth]{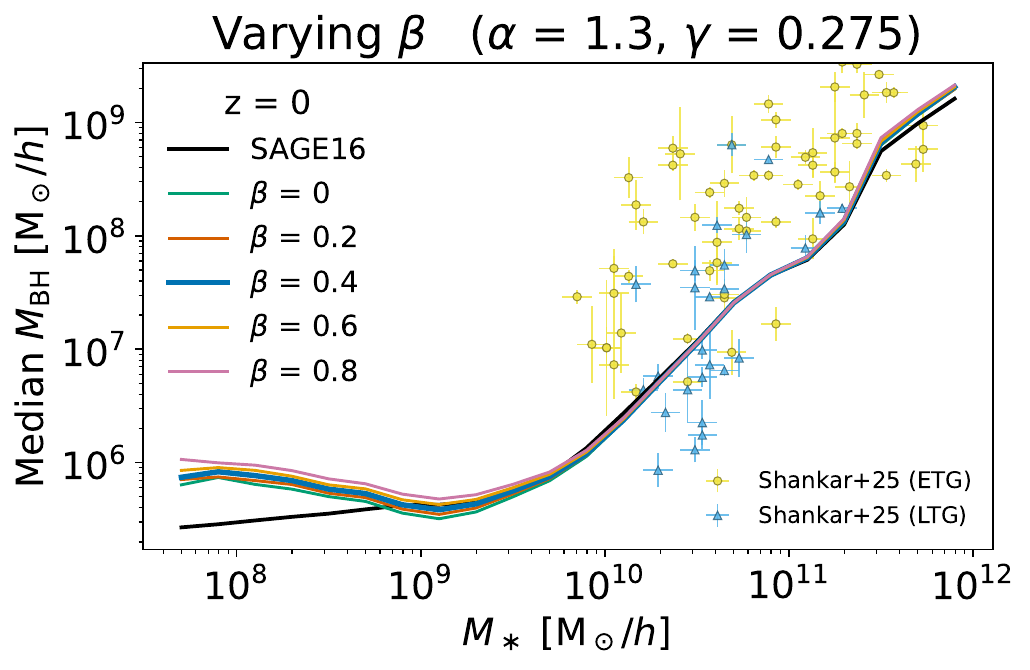}\\
\includegraphics[width=\columnwidth]{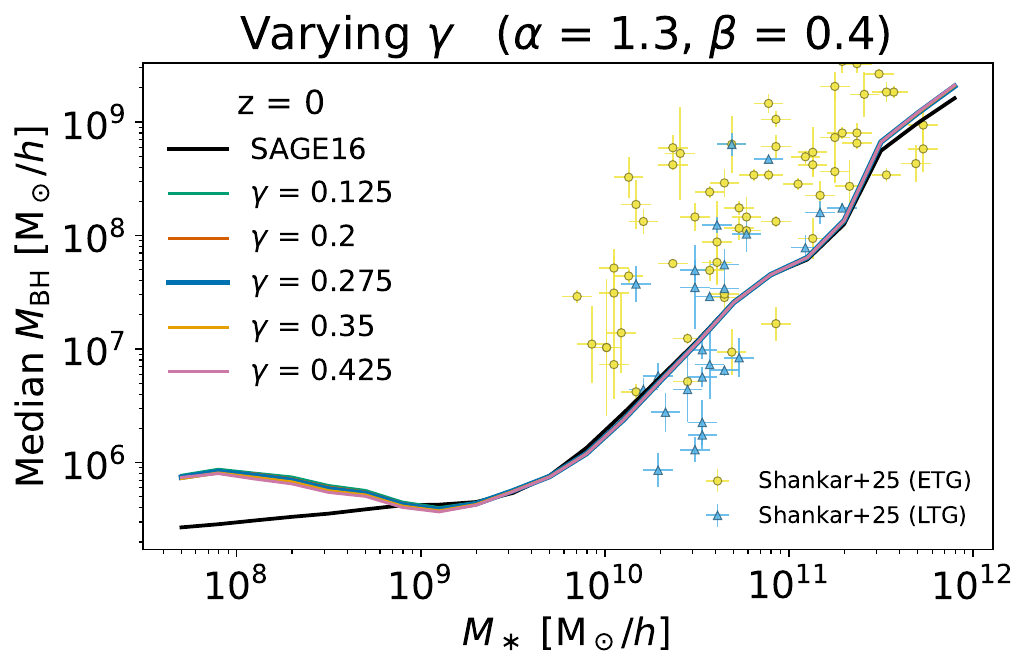}
\caption{The median stellar-black hole mass relation under the same one-parameter variations as \cref{fig:sweep_seedspec} (top to bottom: $\check{\alpha}$, $\check{\beta}$, $\check{\gamma}$), with the stock \textsc{Sage} relation in black and the optimised configuration of \cref{tab:auroraparams} in bold in every panel. Between $10^{9}$ and $10^{11}\,\text{M}_\odot/h$ the relation is insensitive to the seeding parameters across their full explored range, with every configuration lying within $0.17$ dex of the calibrated one; only the strongest overdensity coupling moves it further, by up to $0.45$ dex at the extremes of stellar mass. Overlaid are the local galaxies with dynamically measured black hole masses compiled by \citet{Shankar2025}, separated into early types (yellow) and late types (sky blue), with their total stellar masses derived from the tabulated $3.6\,\mu$m magnitudes using the mass-to-light ratio those authors adopt. The data are biased high in stellar velocity dispersion at fixed luminosity, constituting an upper envelope rather than an unbiased mean.}
\label{fig:sweep_smbh}
\end{figure}

We now vary the three coupling indices one at a time about an anchor configuration $(\check{\alpha}, \check{\beta}, \check{\gamma}) = (1.3,\, 0.4,\, 0.275)$, which is the calibrated configuration of \cref{tab:auroraparams}. \Cref{fig:sweep_seedspec} shows the resulting seed mass spectra $(\psi)$, while \cref{fig:sweep_bhmf} shows the $z=0$ black hole mass functions $(\phi)$ against the stock model, and \cref{fig:sweep_smbh} shows the $z=0$ stellar-black hole mass relation. In each figure the optimised configuration of \cref{tab:auroraparams} is drawn in bold, with the remaining curves bracketing it from below and above in the varied index.

Varying $\check{\alpha}$ visibly distorts the seed mass spectrum: it is very narrow when the coupling is weak and broad when it is strong (\cref{fig:sweep_seedspec}, top panel). This behaviour follows directly from the term's definition: $\check{\alpha}$ measures the coupling of the seed mass to the cosmic variance of the density field, and strengthening it preferentially seeds massive black holes in dense environments while suppressing those in voids, stretching the spectrum across its full dynamic range. The imprint survives to $z=0$ as a corresponding broadening of the black hole mass function (\cref{fig:sweep_bhmf}), where the stock model's narrow seed injection is replaced by an environmentally graded one.

Varying $\check{\beta}$ acts differently: it elevates the spectrum without profoundly affecting its shape (\cref{fig:sweep_seedspec}, middle panel). The radial skewness selects the most concentrated configurations, the most interaction-prone and accretion-prone seeding conditions, and it does so largely independently of halo mass and of the cosmic variance traced by $\delta$. In other words, a strongly asymmetric neighbourhood can occur at any overdensity. Strengthening the coupling therefore lifts most seeds slightly and the most asymmetric substantially, rather than re-grading the spectrum between environments, and its signature at $z=0$ is confined to the abundance of the most massive black holes.

Varying $\check{\gamma}$ leaves the bulk of the seed mass spectrum unchanged but separates its high mass tail monotonically (\cref{fig:sweep_seedspec}, bottom panel). The asymmetry is structural. Seeds are assigned at the moment a lineage first crosses the eligibility threshold, so the mass ratio entering the halo mass term of \Cref{eq:BHseeding} is close to unity for almost every seeding event, and the term is appreciable only for haloes that cross with substantial overshoot: the rapidly growing systems that gain many particles within a single snapshot interval.

The typical halo crosses the threshold with almost no overshoot, so the bulk of the spectrum, and with it the bulk of the $z=0$ black hole mass function (\cref{fig:sweep_bhmf}), is left where the other two couplings put it. The overshooting haloes the $\check{\gamma}$ term selects are instead those responsible for the diffuse cloud above the main relation of \cref{fig:seedmassspec}. What \textsc{Millennium} establishes is that the coupling acts at all; what it cannot establish is how much it is worth, since giving it real dynamic range is a question of mass resolution rather than of volume. We quantify its reach against the other two in \Cref{sec:couplings}, and return to the resolution question in \Cref{sec:disc_extend}.

Throughout these variations, the median stellar-black hole mass relation is left essentially unchanged between $10^{9}$ and $10^{11}\,\text{M}_\odot/h$ (\cref{fig:sweep_smbh}), as is the stellar-to-halo mass relation (not shown), though the scatter about the former does respond at high redshift, as shown in \Cref{sec:pathway}. The insensitivity is expected, given that seed masses lie well below the self-regulated masses that \textsc{Sage}'s accretion physics establishes at fixed stellar mass, meaning that the relation is growth-dominated. For the same reason, no seeding prescription can close the offset, of up to two orders of magnitude, between the model medians and the upper envelope of dynamically measured early types in \cref{fig:sweep_smbh}. The offset is common to every configuration, the stock model included, and owes to the accretion physics of the base model and the selection of the dynamical sample rather than to seeding. However, this is also the property that makes \textsc{Aurora} practically useful: the seeding prescription can be tested and varied flexibly, across its full parameter range, without any detriment to the galaxy statistics against which the base model is calibrated.

\section{The Growth of Massive Black Holes and Quiescent Galaxies}
\label{sec:evolution}

\Cref{sec:optimisation} established how the three coupling indices shape the seed mass spectrum $\psi$ and the $z=0$ black hole mass function $\phi$. We now follow their consequences forward in time: how the environmentally graded seeds of \Cref{eq:BHseeding} are assembled into massive black holes, and what imprint that assembly leaves on the quiescent galaxy population. Throughout, results are compared against the stock \textsc{Sage} run on identical merger trees, and a galaxy is classified as quiescent when $\text{sSFR} < 0.2\,/\,t(z)$, with $t(z)$ the age of the universe at the redshift of the snapshot, evaluated in the simulation cosmology. This is the threshold of \citet{Pacifici2016}, as applied to massive quiescent galaxies at $3 < z < 5$ by \citet{Carnallmnras}.

Six quantities, defined formally in \cref{tab:metrics}, summarise each run. Two describe the black hole mass budget of the whole volume: the seed budget fraction $\eta$, the share of all black hole mass ever formed which was delivered as seeds rather than accreted afterwards; and the black hole mass amplification $\Xi$, the total black hole mass at $z=0$ relative to that of the calibrated configuration. The seed retention $\zeta$ describes individual lineages, as the median fraction of a galaxy's black hole mass still contributed by its own seed; not including accreted seeds as $\eta$ does. The remaining three describe quenching among the seeded galaxies of haloes above $10^{12}\,\text{M}_\odot/h$. They include the quiescent enhancement $\Theta$, the amount by which the quiescent fraction $f_q$ of a configuration exceeds that of the stock \textsc{Sage} model; $\Theta_i$, the same quantity within the $i$th tercile of the seeding environment, numbered from the lowest values ($i=1$) to the highest ($i=3$); and the environmental gradient $\Delta\Theta$ between the outer terciles, defined $\Delta\Theta \equiv \Theta_3 - \Theta_1$.

We divide into terciles because three classes are what the analysis requires: sparse, average and dense environments. Since $\delta$ is normalised to the geometric mean density over all haloes (\Cref{sec:aurora_metrics}), the middle tercile brackets that reference density, $\delta = 0$, leaving the outer two to isolate the underdense and overdense regimes in which the overdensity coupling acts in opposite directions (\Cref{sec:couplings}). The same division separates the least, moderately and most asymmetric neighbourhoods in $\mu_3$, and the lowest, middle and highest seeding halo masses. The boundaries are fixed once rather than redrawn at each snapshot, so that each class keeps the same meaning at every epoch, at the cost of unequal occupancy at any one (\cref{tab:metrics}). The environmental gradient $\Delta\Theta$ then measures how much more strongly a configuration enhances quenching in the highest class than in the lowest, or, for the overdensity, in dense environments rather than in sparse ones.

\Cref{tab:metrics} gives these quantities for six runs: the stock \textsc{Sage16} model; the calibrated \textsc{Aurora} configuration, $(\check{\alpha}, \check{\beta}, \check{\gamma}) = (1.3,\, 0.4,\, 0.275)$; three configurations in which a single index is raised to the top of its sweep with the other two held at their calibrated values, $(2.6,\, 0.4,\, 0.275)$, $(1.3,\, 0.8,\, 0.275)$ and $(1.3,\, 0.4,\, 0.425)$, each drawn in pink in our figures; and a control.

In the control, the calibrated model is rerun with the environment of every seeding halo replaced by a value drawn from the environment distribution itself, by the quantile mapping of \citet{Chittenden}. The replacement is drawn within fixed-snapshot bins of halo mass $0.1$ dex wide, and from the seeding events alone. The seed mass spectrum, its dependence on halo mass and epoch, the haloes which seed and the epochs at which they do so are thus all preserved; what is destroyed is the correspondence between a halo and the density field in which it formed. We randomise $\delta$ and $\mu_3$ independently rather than as a pair, which in this volume is close to the same operation, since their rank correlation across seeding events is $-0.038$ (\Cref{sec:couplings}). Trees, merger histories and every parameter of \cref{tab:auroraparams} are unchanged. Any quantity the prescription sets through environment must fall towards the stock model under this control, and any quantity it sets through the seed mass distribution alone must survive. It is the most direct test in this work of whether a signal is environmental, and \cref{tab:metrics} applies it to every quiescence statistic.

\begin{table*}
\caption{Quantitative, dimensionless metrics used throughout \Cref{sec:evolution} to compare \textsc{Aurora} against the stock \textsc{Sage} model. The upper block defines each metric; the lower block gives its value for the stock model, for the calibrated configuration of \cref{tab:auroraparams}, $(\check{\alpha}, \check{\beta}, \check{\gamma}) = (1.3,\, 0.4,\, 0.275)$, for each coupling index raised to the upper end of its sweep with the other two held at their calibrated values, and for the calibrated configuration rerun with scrambled environments (\Cref{sec:evolution}). In these definitions, $M_\bullet$ is the seed mass assigned by \textsc{Aurora} (for $\zeta$, the seed planted in the galaxy's own main branch), $M_\text{BH}$ is the black hole mass, and $f_q$ is the fraction of seeded galaxies in haloes above $10^{12}\,\text{M}_\odot/h$, satellites and orphans included, whose specific star formation rate lies below the threshold of $0.2/t(z)$. Sums over 'events' run over every seeding event in the volume across all snapshots, while sums over 'acc' cover every accretion increment onto any black hole. $\zeta$, $\Theta$ and the $\Theta_i$ are quoted at the $z = 3.06$ snapshot, where the haloes above $10^{12}\,\text{M}_\odot/h$ hold $197\,727$ seeded galaxies, the same in every run to within on. $\zeta$ is the median over the $87\,961$ centrals of these haloes. $\eta$ and $\Xi$ are cumulative to $z=0$. The 'split' column names the seeding epoch quantity whose terciles define $\Theta_1$, $\Theta_2$ and $\Theta_3$, numbered from its lowest values to its highest: each configuration is divided by the quantity its varied index couples to, as in \crefrange{fig:fqenv_alpha}{fig:fqenv_gamma}, with the calibrated configuration given in all three since it is the reference curve of every panel, and the scrambled control likewise. Tercile boundaries are fixed once from the massive central population pooled over all snapshots, at $\delta = -0.31$ and $0.35$, $\mu_3 = 0.54$ and $1.14$, and $\log_{10} M_h^\text{seed} = 10.92$ and $11.01$. Being pooled, they divide any single snapshot unevenly: at $z = 3.06$ the overdensity boundaries place 7.9, 10 and 82 per cent of the seeded galaxies in the sparsest, middle and densest terciles, while the other two splits keep every tercile between 32 and 36 per cent. Metrics defined relative to the stock model are identically zero for it by construction, and $\eta$ and $\zeta$ are undefined there, as the stock prescription has no explicit seed. Raising $\check{\gamma}$ to the top of its sweep changes neither $f_q$, $\Theta$ nor any $\Theta_i$ by more than $0.007$; its response is confined to the mass budget, raising $\eta$ by $14$ per cent and $\Xi$ by $7.3$ per cent.}
\label{tab:metrics}
\begin{tabular}{cllr}
\hline
Symbol & Metric & Definition & \\
\hline
$\eta$ & Seed budget fraction &
  $\displaystyle \eta = \frac{\sum_\text{events} M_\bullet}
  {\sum_\text{events} M_\bullet + \sum_\text{acc} \Delta M_\text{BH}}$ &
  \\[2ex]
$\zeta$ & Seed retention &
  $\displaystyle \zeta = \text{med}\left(\frac{M_\bullet}{M_\text{BH}}\right)$ &
  \\[2ex]
$\Xi$ & Black hole mass amplification &
  $\displaystyle \Xi = \frac{\sum_{z=0} M_\text{BH}}{\sum_{z=0} M_\text{BH}^\text{cal}}$ &
  \\[2ex]
$\Theta$ & Quiescent enhancement &
  $\displaystyle \Theta = f_q - f_q^{\textsc{Sage}}$ &
  \\[1ex]
$\Theta_i$ & Quiescent enhancement, tercile $i$ &
  $\displaystyle \Theta_i = f_{q,i} - f_{q,i}^{\textsc{Sage}}$ &
  \\[1ex]
$\Delta\Theta$ & Environmental gradient &
  $\displaystyle \Delta\Theta = \Theta_3 - \Theta_1$ &
  \\
\hline
\end{tabular}

\vspace{2ex}

\begin{tabular}{llcccccccccc}
\hline
Configuration & $(\check{\alpha},\, \check{\beta},\, \check{\gamma})$ & $\eta$ & $\zeta$ & $\Xi$ & $f_q$ & $\Theta$ & Split &
  $\Theta_1$ & $\Theta_2$ & $\Theta_3$ & $\Delta\Theta$\\
\hline
Stock \textsc{Sage16} & ---
  & --- & --- & $0.852$ & $0.40$ & $\equiv 0$ & --- & $\equiv 0$ & $\equiv 0$ & $\equiv 0$ & $\equiv 0$\\
\hline
Calibrated & $(1.3,\, 0.4,\, 0.275)$
  & $0.29$ & $0.084$ & $\equiv 1$ & $0.48$ & $+0.082$
  & $\delta$ & $-0.028$ & $-0.012$ & $+0.104$ & $+0.132$\\
  & & & & & & & $\mu_3$ & $+0.042$ & $+0.085$ & $+0.114$ & $+0.072$\\
  & & & & & & & $M_h$ & $+0.098$ & $+0.052$ & $+0.092$ & $-0.006$\\
\hline
$\check{\alpha}$ raised & $(2.6,\, 0.4,\, 0.275)$
  & $0.74$ & $0.169$ & $1.867$ & $0.68$ & $+0.280$
  & $\delta$ & $-0.015$ & $+0.009$ & $+0.341$ & $+0.357$\\
$\check{\beta}$ raised & $(1.3,\, 0.8,\, 0.275)$
  & $0.37$ & $0.124$ & $1.050$ & $0.53$ & $+0.135$
  & $\mu_3$ & $+0.051$ & $+0.130$ & $+0.215$ & $+0.164$\\
$\check{\gamma}$ raised & $(1.3,\, 0.4,\, 0.425)$
  & $0.33$ & $0.083$ & $1.073$ & $0.48$ & $+0.080$
  & $M_h$ & $+0.091$ & $+0.051$ & $+0.097$ & $+0.006$\\
\hline
Scrambled environments & $(1.3,\, 0.4,\, 0.275)$
  & $0.26$ & $0.060$ & $1.003$ & $0.40$ & $+0.002$
  & $\delta$ & $+0.009$ & $+0.005$ & $+0.001$ & $-0.008$\\
  & & & & & & & $\mu_3$ & $+0.005$ & $+0.004$ & $-0.002$ & $-0.008$ \\
  & & & & & & & $M_h$ & $+0.014$ & $-0.012$ & $+0.002$ & $-0.012$ \\
\hline
\end{tabular}
\end{table*}

\subsection{The structure of the three couplings}
\label{sec:couplings}

The three terms of \Cref{eq:BHseeding} do not act in the same way, and the difference follows from the distributions of the quantities they multiply rather than from the couplings themselves. Over the 4.6 million seeding events in \textsc{Millennium}, the overdensity term $\ln(1+\delta)$ has a median of $+0.002$, with 50 per cent of events positive; because $\delta$ is normalised to the geometric mean density (\Cref{sec:aurora_metrics}), the term is centred on zero by construction. Raising $\check{\alpha}$ therefore \emph{pivots} the seed spectrum, boosting seeds in overdense environments and suppressing them by a comparable factor in voids. The skew term behaves quite differently: $\mu_3$ has a median of $+0.44$ and is positive for 72 per cent of seeding events, so raising $\check{\beta}$ lifts the seed mass of the great majority of the population, most strongly where the neighbourhood is most asymmetric. The two terms are also close to orthogonal, with a Spearman rank correlation between $\delta$ and $\mu_3$ of only $-0.038$ across seeding events.

This is the quantitative form of the distinction drawn in \Cref{sec:sweeps}: $\check{\alpha}$ measures the coupling to cosmic variance, which is symmetric about the mean density and therefore redistributive, whereas $\check{\beta}$ measures the coupling to interaction-prone configurations, which are the norm rather than the exception and therefore produce a one-sided enhancement.

The halo mass term behaves differently again, for a reason specific to the eligibility criterion rather than to the coupling. Seeds are assigned at the moment a lineage first crosses $\check{N}_h^\text{min}$, so the mass ratio entering the third term of \Cref{eq:BHseeding} is close to unity for almost every event. The term is therefore noticeable only on haloes crossing the threshold with substantial overshoot, which are a small fraction of all crossings; but while the fraction is set by the particle mass, the number is set by the volume, and \textsc{Millennium} supplies enough of them for the coupling to register (\Cref{sec:sweeps}).

\Cref{tab:metrics} scales the reach of the three couplings against one another. Each comparison starts from the calibrated configuration, which reproduces the stock black hole population at $z=0$ (\Cref{sec:calibration}), and so measures how far each coupling can move a calibrated model. The top of each sweep doubles $\check{\alpha}$ and $\check{\beta}$, but raises $\check{\gamma}$ by only 55 per cent. Over those steps, the total seed mass injected in the volume grows by factors of $6.7$, $1.43$ and $1.21$ respectively, and the seed budget fraction rises from $\eta = 0.29$ to $0.74$, $0.37$ and $0.33$. The run with $\check{\gamma}$ raised to $0.425$ shows that the halo mass coupling acts on the mass budget alone: neither $f_q$, $\Theta$ nor any $\Theta_i$ moves by more than $0.007$ from its calibrated value, yet the black hole mass amplification rises to $\Xi = 1.073$, above the $1.050$ reached with $\check{\beta}$ raised to $0.8$, despite the smaller step. Of the three couplings, the halo mass coupling is therefore the only one whose signature is confined to the mass budget and the high mass tail of the seed spectrum, rather than shaping the galaxy population as a whole.

\subsection{Assembly of massive black holes from seeds}
\label{sec:synthesis}

With \textsc{Aurora} as the sole seeding channel, every black hole descends from a seed placed by \Cref{eq:BHseeding}, and the mass budget can be audited exactly, which we do through the quantities $\eta$, $\Xi$ and $\zeta$ of \cref{tab:metrics}.

The seed budget fraction $\eta$ is the share of all black hole mass ever formed in the volume which was delivered as seed mass rather than accreted afterwards. The calibrated configuration gives $\eta = 0.29$, i.e. seeding supplies 29 per cent of the black hole mass ever formed, with quasar-mode, radio-mode and post-seeding Eddington accretion constituting the remainder. Strengthening the environmental couplings raises $\eta$ to $0.37$ at $\check{\beta} = 0.8$, and to $0.74$ at $\check{\alpha} = 2.6$, at which point seeding rather than accretion supplies the majority of the black hole mass budget.

The black hole mass amplification relative to the calibrated model is denoted $\Xi$, and refers to the surviving population, and is therefore the quantity a $z=0$ observer would measure. It reaches $\Xi = 1.87$ over the same range as above. For reference, the stock model sits at $\Xi = 0.85$, i.e. gating black hole formation on the \textsc{Aurora} eligibility criterion and seeding it environmentally yields $17$ per cent more black hole mass by $z=0$ than the merger-driven prescription it replaces, even at the calibrated coupling.

\begin{figure}
\includegraphics[width=\columnwidth]{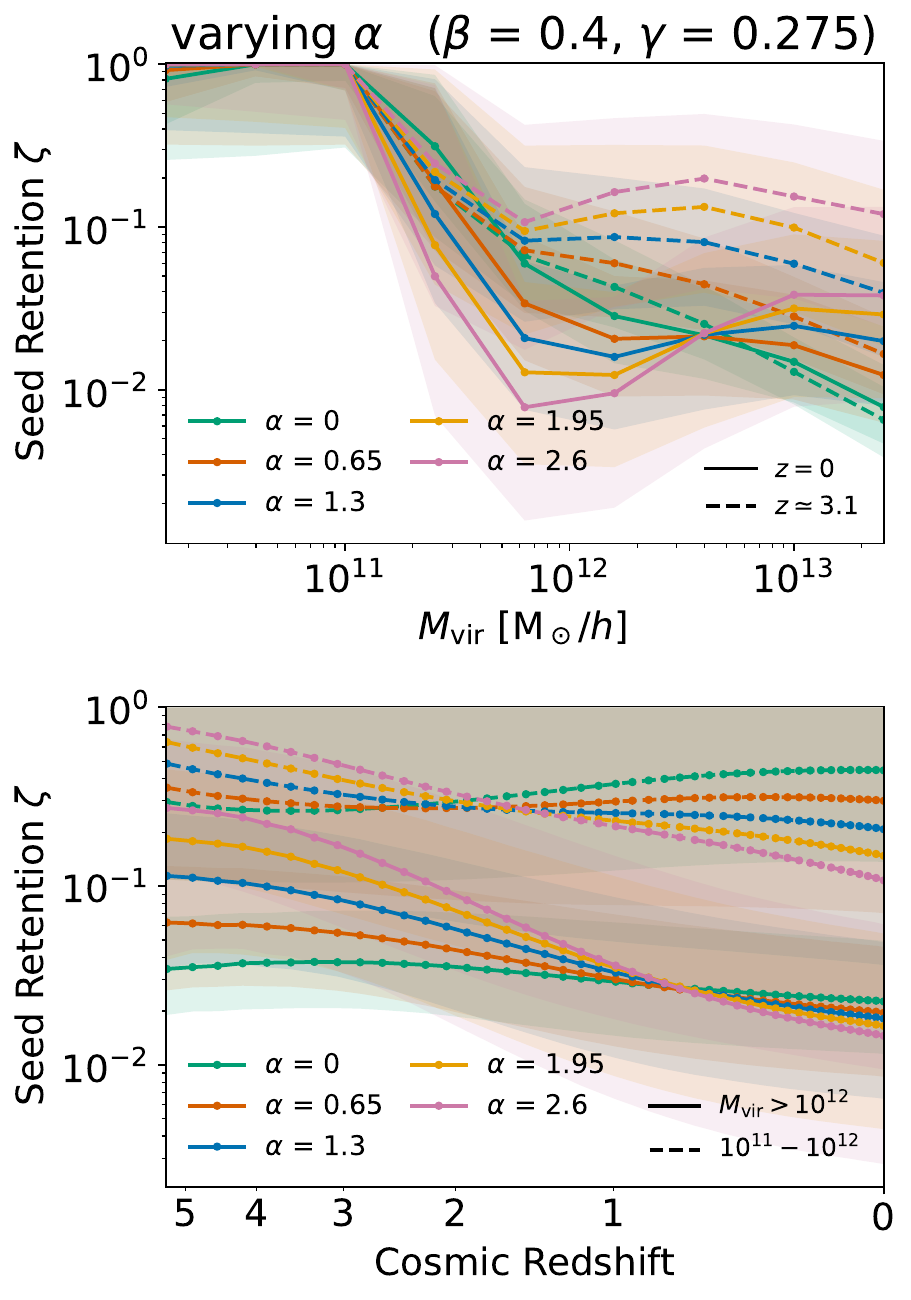}
\caption{Seed retention $\zeta = M_\bullet/M_\text{BH}$, the fraction of a galaxy's black hole mass still contributed by the seed planted in its own lineage, for central galaxies in the calibrated configuration and with varying $\check{\alpha}$. Lines are medians and shaded bands show the interquartile range. \emph{Upper:} against halo mass, at $z = 0$ (solid) and $z \simeq 3.1$ (dashed); retention is complete in the lowest mass haloes and falls by nearly two orders of magnitude across the resolved range, and is higher at earlier times at fixed mass. \emph{Lower:} against redshift, for haloes above $10^{12}$ (solid) and between $10^{11}$ and $10^{12}\,\text{M}_\odot/h$ (dashed). At $z \simeq 3.1$, strengthening $\check{\alpha}$ raises $\zeta$ at every halo mass above $10^{11}\,\text{M}_\odot/h$, since heavier seeds constitute a larger fraction of the black holes they grow into, yet it declines more rapidly as more heavy seeds are accreted: by $z = 0$ the ordering has inverted in haloes below $10^{13}\,\text{M}_\odot/h$, and survives only above that mass.}
\label{fig:zetahalo}
\end{figure}

It is useful to distinguish this population level budget from the fate of an individual lineage. The seed retention $\zeta$ of \cref{tab:metrics} is the median ratio of a galaxy's own seed mass to its current black hole mass: unity for a black hole which has not grown since seeding, and small for one whose mass is dominated by accretion and by the black holes of merged companions. \Cref{fig:zetahalo} shows this behaviour for $\check{\alpha}$, while $\check{\beta}$, not shown, behaves similarly, without the inversion described below. Retention is essentially unity in haloes below $10^{11}\,\text{M}_\odot/h$ and falls to $0.018$ above $10^{12}\,\text{M}_\odot/h$ at $z=0$. It is, however, a strong function of redshift, since less assembly has intervened at earlier times: above $10^{12}\,\text{M}_\odot/h$ it is $\zeta = 0.084$ at $z \simeq 3.1$ and $0.10$ at $z \simeq 4.2$. At $z \simeq 3.1$, strengthening either active coupling raises it in these haloes, to $\zeta = 0.17$ at $\check{\alpha} = 2.6$ and $0.12$ at $\check{\beta} = 0.8$ (\cref{tab:metrics}), since heavier seeds remain a larger fraction of the black holes they grow into. For $\check{\alpha}$ this relationship inverts at lower redshifts in haloes below $10^{13}\,\text{M}_\odot/h$ (\cref{fig:zetahalo}), likely due to the onset of accretion: at $z = 0$ the median over all haloes above $10^{12}\,\text{M}_\odot/h$, most of which lie below $10^{13}$, is $0.015$ at $\check{\alpha} = 2.6$, against $0.018$ in the calibrated configuration. Above $10^{13}\,\text{M}_\odot/h$ the ordering holds at every redshift, as it does at every mass for $\check{\beta}$.

\subsubsection{A seed-dominated growth pathway}
\label{sec:pathway}

\begin{figure}
\includegraphics[width=\columnwidth]{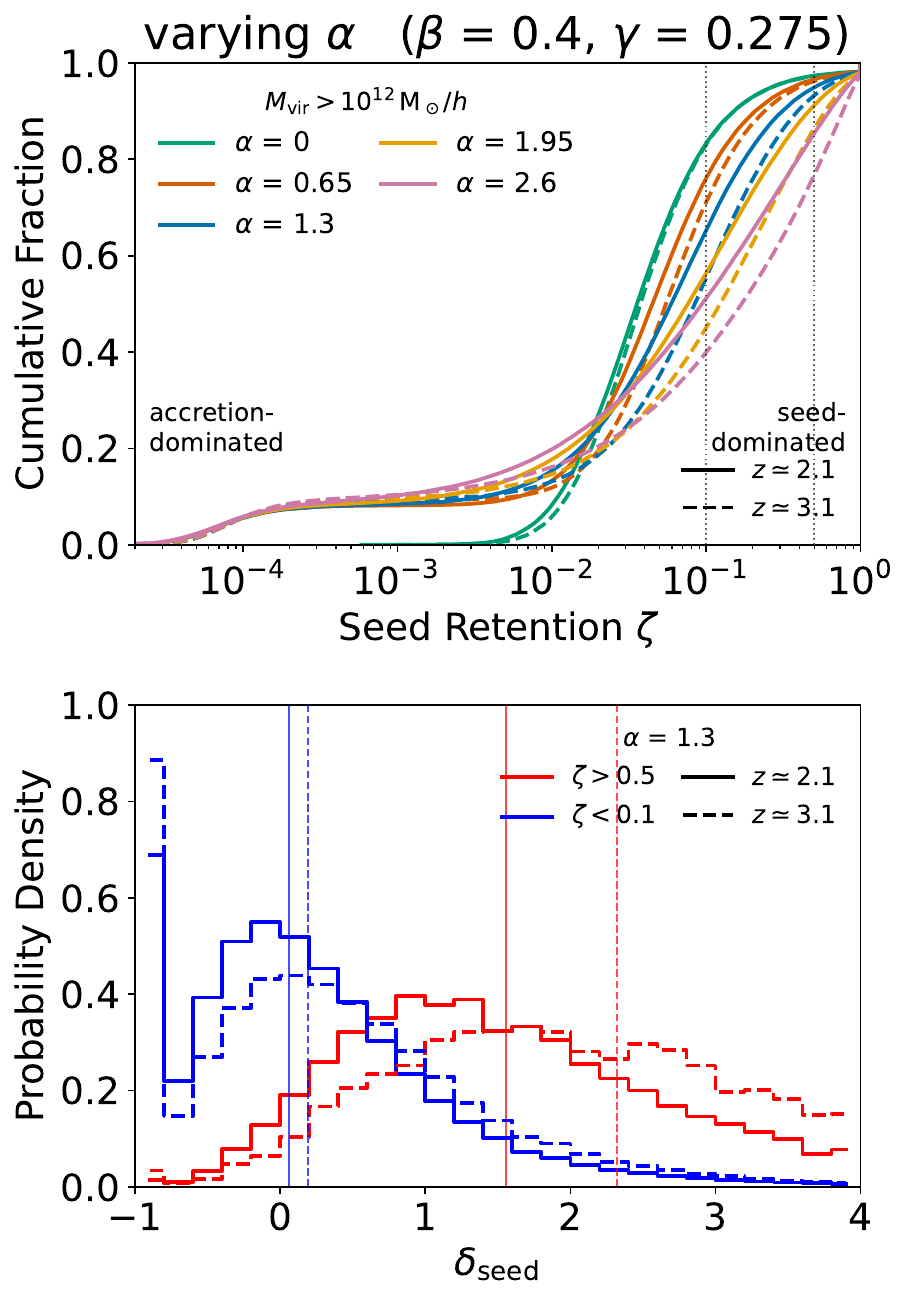}
\caption{The seed-dominated growth pathway. \emph{Upper:} cumulative distribution of seed retention among central galaxies with haloes above $10^{12}\,\text{M}_\odot/h$, at $z \simeq 2.1$ (solid) and $z \simeq 3.1$ (dashed), for the same configurations as \cref{fig:zetahalo}. Dotted lines mark the $\zeta < 0.1$ and $\zeta > 0.5$ regimes discussed in the text; the seed-dominated tail is a minority at every coupling, but grows with the strength of the environmental couplings and towards higher redshift. \emph{Lower:} distribution of the seeding epoch overdensity for the two regimes in the calibrated configuration, with medians marked. Systems which retain their seeds were seeded in markedly denser environments than the accretion-dominated majority, and the separation widens with redshift.}
\label{fig:zetapathway}
\end{figure}

Retention is not uniform among massive haloes, and its distribution identifies a physically distinct subpopulation of central galaxies (\cref{fig:zetapathway}). Among the centrals of haloes above $10^{12}\,\text{M}_\odot/h$ at $z \simeq 2.1$, a minority of 5.1 per cent retain more than half of their black hole mass as their original seed (6.8 per cent at $z \simeq 3.1$), while the majority retain less than 10 per cent. The two groups differ sharply in their seeding environments: the seed-dominated centrals were seeded at a median overdensity of $\delta = +1.6$ and skew $\mu_3 = +1.2$, against $+0.06$ and $+0.81$ for the accretion-dominated majority. At $z \simeq 3.1$ the contrast is starker still, at $\delta = +2.3$ against $+0.19$. They also respond quite differently to the seeding parameters. Raising $\check{\alpha}$ to $2.6$ changes the black hole masses of the seed-dominated centrals by $+0.40$ dex at $z \simeq 2.1$ and $+0.55$ dex at $z \simeq 3.1$, and raising $\check{\beta}$ to $0.8$ changes them by $+0.18$ and $+0.11$ dex respectively, whereas the accretion-dominated majority do not shift noticeably under either variation.

This subpopulation is the physically interesting one. A massive galaxy whose black hole is still dominated by its seed is one which grew rapidly, in a dense and asymmetric environment, without the extensive assembly history which would otherwise dilute the seed among many merged companions. Such systems inherit the environmental grading of \Cref{eq:BHseeding} essentially intact, and carry it into their feedback history. They resemble the channel identified by \citet{Chittenden3} in \textsc{Thesan}, in which the earliest \acp{mqg} assemble through smooth accretion in dense environments and show little influence from mergers, and in which the radial skewness of \citet{Chittenden} was found to trace the anisotropic inflows responsible for their growth. The black holes of this channel are overmassive with respect to their hosts, but their growth is affected by an implementation error in the \ac{agn} feedback of the public \textsc{Thesan} release;\footnote{The black hole routines of the public \textsc{Thesan} release adopt the reduced speed of light of its radiative transfer, cutting the energy injected by both quasar- and radio-mode \ac{agn} feedback by a factor of 25 and permitting accretion at up to five times the Eddington limit \citep{Garaldi2026}. \citet{Chittenden3} treat the affected runs as a limiting case of weakly regulated growth, and show the halo and environmental conditions of the channel, the only information \textsc{Aurora} reads, to be insensitive to the subgrid model.} the correspondence therefore holds for the conditions in which the channel assembles, not for the black hole masses it produces.

\textsc{Colibre} reaches the same association by a different route: its \acp{mqg} seed their black holes slightly earlier, in slightly more massive haloes, and grow them far faster from $z \simeq 8$. They occupy systematically more overdense regions than star forming galaxies of equal mass at $z \gtrsim 4$, on local ($0.3$ cMpc) and cosmic web ($1.0$ cMpc) scales, with the contrast washing out by the selection epoch \citep{ChandroGomez2}. The two simulations differ on the role of mergers: \textsc{Colibre} attributes part of the accelerated assembly in dense regions to a higher merger rate, whereas \textsc{Thesan}'s \acp{mqg} are merger-poor. \textsc{Aurora} is neutral on that point: $59$ per cent of the seed-dominated centrals have had no major merger, against $56$ per cent of the accretion-dominated majority. What both simulations share, and what \textsc{Aurora} parameterises, is the earlier and faster growth of black holes in denser environments. \textsc{Aurora} produces a comparable pathway from dark matter information alone, and makes its incidence a function of the seeding parameters rather than of the accretion prescription. Agreement with two hydrodynamical simulations whose subgrid physics differs, and in \textsc{Thesan} includes the implementation error noted above, nonetheless constitutes a check of consistency rather than a test of the physics.

A related consequence is that the black holes whose masses are most strongly altered are not necessarily those in the most massive galaxies: their hosts quench before reaching the stellar mass they attain in the stock model. \textsc{Aurora} thus broadens the galaxy-black hole mass relation, displacing it only below $10^{10}\,\text{M}_\odot$, as a direct consequence of tying the seed to the environment rather than to the halo alone.

\begin{figure}
\includegraphics[width=\columnwidth]{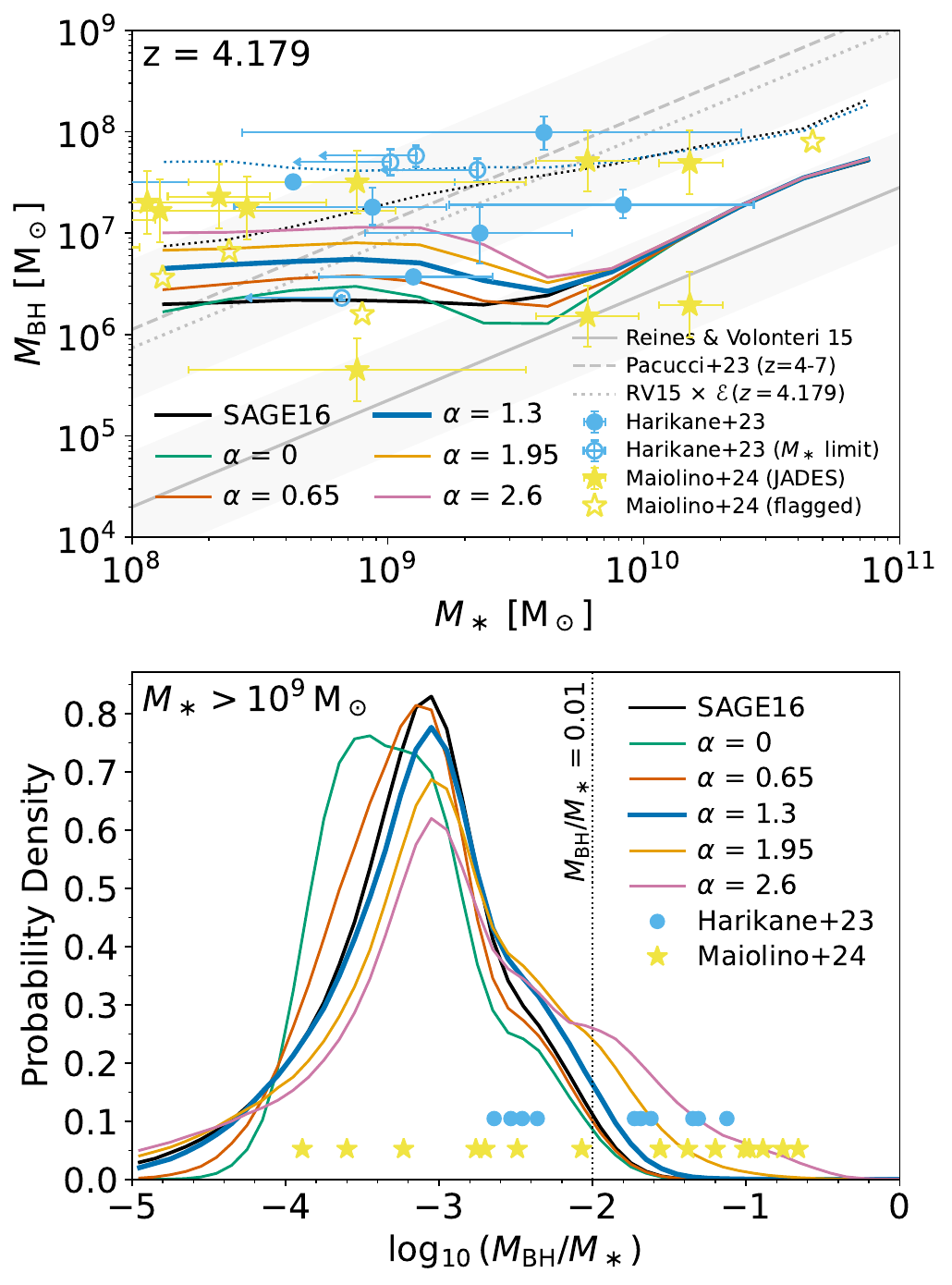}
\caption{The galaxy-black hole mass relation at $z = 4.179$. \emph{Upper:} black hole mass against stellar mass at $z = 4.179$, for the $\check{\alpha}$ sweep and for the stock model. Solid lines indicate medians, and the blue and black dotted lines show the $99$th percentile of the calibrated \textsc{Aurora} and stock \textsc{Sage} models, respectively. Overlaid are the JADES broad-line \acp{agn} of \citet{Maiolino} (yellow), and the broad-line \acp{agn} of \citet{Harikane2023} (sky blue), where open symbols mark the objects whose stellar masses those works flag as uncertain, and leftward arrows mark the four hosts of \citet{Harikane2023} whose stellar masses are upper limits, and whose mass ratios are therefore lower bounds. Three straight lines are drawn in grey. The local ($z = 0$) relation of \citet{ReinesVolonteri} (solid) and the same relation scaled by the redshift evolution function $\mathscr{E}(z)$ of \citet{PacucciLoeb} (dotted) are shown for reference, whereas the $z = 4-7$ relation of \citet{Pacucci} (dashed) is a fit to the observed hosts themselves, with $20$ of its $21$ hosts drawn from the \citet{Harikane2023} and \citet{Maiolino} samples shown here. Shaded regions show the $16-84$th percentile margins of these three relations. \emph{Lower:} distribution of $\log_{10}(M_\text{BH}/M_*)$ for galaxies above $10^{9}\,\text{M}_\odot$, for the $\check{\alpha}$ sweep and the stock model, with the individual \citet{Harikane2023} and \citet{Maiolino} objects marked along the axis and $M_\text{BH}/M_* = 10^{-2}$ indicated. \textsc{Aurora} leaves the bulk of the distribution where the stock model puts it and extends its high ratio tail, but that tail still falls short of most of the observed objects. The observed sample is, moreover, biased towards overmassive systems by selection \citep{Li2025}, and its virial black hole masses may themselves be overestimated (\citealt{Rusakov2026}; but see \citealt{Brazzini2025}).}
\label{fig:mbhmstar}
\end{figure}

\Cref{fig:mbhmstar} shows the resulting $M_\text{BH}-M_*$ plane at $z=4.179$, an epoch at which black holes overmassive relative to their hosts are reported in abundance \citep{Harikane2023, Maiolino, Ubler, Kocevski2025}, many of them \acp{lrd}. Above $10^{10}\,\text{M}_\odot$ the median relation is indistinguishable from that of the stock model, agreeing to within $0.03$ dex in $\log_{10}(M_\text{BH}/M_*)$; below $10^{9.5}\,\text{M}_\odot$ it is raised by up to $0.40$ dex. Below $10^{10}\,\text{M}_\odot$, where every observed object above the stock model's $99$th percentile lies, the high ratio tail is extended at every mass. Among the galaxies above $10^{9}\,\text{M}_\odot$, the median ratio moves by only $+0.05$ dex, while the fraction exceeding $10^{-2}$ rises from $1.9$ to $3.7$ per cent, the fraction exceeding $3 \times 10^{-2}$ rises from $0.04$ to $0.22$ per cent, and a population above $10^{-1}$ appears where the stock model has almost none.

Over the galaxies above $10^{9}\,\text{M}_\odot$ as a whole, the prescription changes the scatter of the relation, not its normalisation, and the broader scatter is consistent with contemporary reionisation-era population statistics. Across nine quasars at $z \simeq 6.2-6.4$, \citet{Silverman} forward-model the selection and recover an intrinsic dispersion of $0.80^{+0.28}_{-0.23}$ dex about a non-evolving mass relation, against the $0.3-0.4$ dex of the local population. Since their inference is of the intrinsic dispersion of the underlying population, it is directly comparable to ours. Measured over the same stellar mass range at $z = 6.197$, \textsc{Aurora} returns a scatter of $0.90$ dex, within the measured interval, while the stock model returns $0.53$ dex, $1.2\sigma$ below the measured value. With nine quasars, the measurement does not yet discriminate decisively between the two, though it favours the broader relation.

Though \textsc{Aurora} generates some black holes overmassive relative to their hosts, it does not reproduce the full observed population. Across all twenty-two broad-line hosts of \citet{Maiolino} and \citet{Harikane2023} with measured stellar masses, whatever their stellar mass, the median is $\log_{10}(M_\text{BH}/M_*) = -1.65$, which \textsc{Aurora} reaches only at its $99.5$th percentile; its $99$th percentile falls $0.10$ dex below the typical observed object. Four further hosts of \citet{Harikane2023} have only upper limits on their stellar masses, so their ratios are lower bounds and the observed median can only rise. Furthermore, \citet{Chon} show in their model that super-Eddington growth is necessary to achieve the higher mass ratios of the \ac{lrd} systems observed by \citet{Ubler, Kocevski2025}. Taken at face value, these masses place the observed population beyond the reach of environmental seeding alone, which moves the tail in the right direction without reaching it.

The observed masses are, however, subject to two caveats, which act in the same direction. Much of the evidence for overmassive black holes at these epochs comes from \acp{lrd} and related compact broad-line \acp{agn} \citep{Pacucci, Maiolino, Kocevski2025}, whose emission is not yet understood \citep{PerezGonzalez2024, deGraaff2026}. If their broad Balmer lines are not virial, but broadened instead by electron scattering in dense ionised gas, single-epoch virial estimates overstate their black hole masses by up to two orders of magnitude \citep{Rusakov2026}, although electron scattering is ruled out as the dominant broadening in at least one well-studied object \citep{Brazzini2025}. Selection biases and measurement uncertainties can, moreover, largely account for the offset of the \textit{JWST} broad-line \acp{agn} above the local relation \citep{Li2025}. Both effects place the observed locus above the intrinsic one, so that correcting for either moves it towards the \textsc{Aurora} tail, and the extent to which super-Eddington growth is required in addition to environmental seeding remains open.

\subsection{Environmental dependence of quenching}
\label{sec:quiescence}

\begin{figure}
\includegraphics[width=\columnwidth]{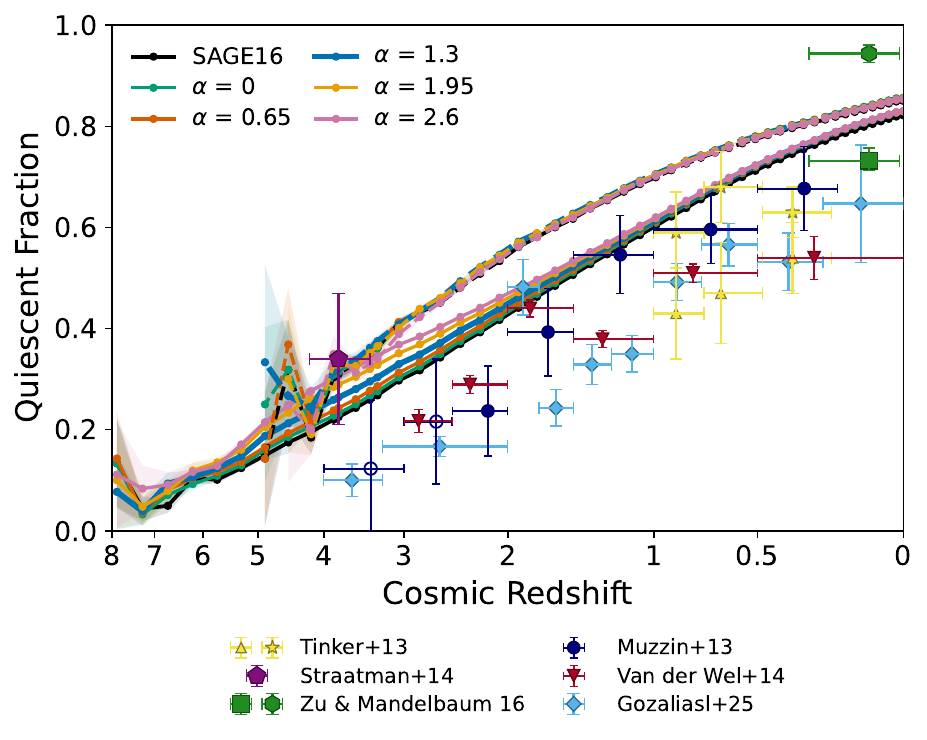}
\caption{Quiescent fraction of all galaxies against redshift, for host haloes above $10^{12}$ (solid) and $10^{13}\,\text{M}_\odot/h$ (dashed), with the stock \textsc{Sage} model in black and the optimised configuration in bold, as $\check{\alpha}$ is varied about the calibrated configuration; the $\check{\beta}$ and $\check{\gamma}$ sweeps are described in the text. Shaded bands are binomial uncertainties. The overdensity coupling shifts the quiescent fraction most strongly at $z \simeq 3-4$. The configurations converge towards $z=0$ without the population becoming saturated, as self-regulation erases the seed dependence of the black hole masses. Published determinations are overlaid, with horizontal bars spanning the redshift interval of each. \citet{Tinker2013}, in yellow, give the quenched fraction of central galaxies at the two halo masses their analysis constrains, $10^{12.3}$ (triangles) and $10^{13.1}\,\text{M}_\odot/h$ (stars). Three determinations select on stellar mass alone, counting every galaxy above $10^{10.6}\,\text{M}_\odot$: \citet{Straatman2014} at $3.4 \le z < 4.2$, in purple; \citet{Muzzin2013}, in navy, from their Schechter fits, with open circles where the fits extrapolate below the completeness limit of the survey; and \citet{vanderWel2014}, in crimson, counted from their sample of early- and late-type galaxies. The dotted curves give the model in the same selection, without the halo selection of the solid curves, for the stock model (black) and the calibrated configuration (blue). \citet{ZuMandelbaum2016}, in forest green, give the red fraction of central galaxies at $z \simeq 0.1$, averaged over all haloes above $10^{12}$ (square) and $10^{13}\,\text{M}_\odot/h$ (hexagon). \citet{Gozaliasl2025}, in sky blue, classify brightest group galaxies. The base model lies above \citet{Tinker2013} and \citet{Gozaliasl2025} at every epoch probed, as their classifications are deliberately conservative. \citet{ZuMandelbaum2016} lie closer, since a classification by colour alone counts dust-reddened star forming galaxies as red. The model exceeds them at the lower threshold and falls short at the higher, because the curves also count satellites, which are less often quiescent than centrals.}
\label{fig:fqhalo}
\end{figure}

The sustained growth of a black hole is supported by the deep potential of its halo \citep{Hopkins}, and since the seeding prescription can influence star formation only through the feedback of a grown black hole, it does so only in haloes massive enough to sustain that growth. Selecting galaxies by halo mass, rather than by the stellar mass which \textsc{Aurora} itself perturbs, isolates this population; \Cref{fig:fqhalo} shows the resulting quiescent fractions, with the stellar mass floor of the surveys imposed for comparison with them. Counting every seeded galaxy in haloes above $10^{12}\,\text{M}_\odot/h$ at $z \simeq 3.1$, the stock model gives $f_q = 0.40$, rising to $0.48$ in the calibrated configuration, $0.53$ at $\check{\beta} = 0.8$ and $0.68$ at $\check{\alpha} = 2.6$, corresponding to enhancements of $\Theta = +0.082$, $+0.14$ and $+0.28$ (\cref{tab:metrics}). The galaxies are common to every run, so these are equivalently excesses of 21, 34 and 70 per cent in the abundance of quiescent galaxies.

The enhancement weakens towards low redshift, with the calibrated configuration falling to 5.8 per cent at $z \simeq 1.1$ and 4.9 per cent at $z = 0$, without the population saturating. The weakening is the signature of self-regulation: as accretion carries each black hole towards the self-regulated mass that \textsc{Sage} sets for its host (\Cref{sec:sweeps}), the memory of the seed which began each lineage is erased (\Cref{sec:synthesis}), taking with it the seeding dependence of the feedback.

Varying $\check{\gamma}$ produces no discernible separation at any redshift, returning $f_q = 0.48$ and $\Theta = +0.08$ at $\check{\gamma} = 0.425$. This is consistent with its behaviour elsewhere rather than at odds with it: the halo mass coupling acts on the high mass tail of the seed spectrum and on the mass budget (\Cref{sec:sweeps}), and neither shifts the quiescent fraction of the population as a whole.

\citet{Tinker2013} constrain the quenched fraction of central galaxies as a function of halo mass over $0.2 < z < 1.0$, and \citet{Gozaliasl2025} classify $\sim\!1700$ brightest group galaxies in \textsc{Cosmos-Web} to $z = 3.7$, in photometric groups whose centrals grow uncertain towards high redshift. The model recovers their redshift dependence but lies above both throughout, by $0.12$ dex on average against \citet{Tinker2013} and $0.26$ dex against \citet{Gozaliasl2025} (\Cref{fig:fqhalo}), as each scheme is built to avoid classifying a star forming galaxy as quiescent. The first displacement is no larger than the difference between the two definitions of \citet{Tinker2013}, $0.1-0.2$ dex, though the second exceeds the differences among the three schemes of \citet{Gozaliasl2025}, up to $0.21$ dex. \citet{ZuMandelbaum2016} instead classify by colour alone, counting dust-reddened star forming galaxies as red; their red fraction at $z \simeq 0.1$, averaged above $10^{13}$ and $10^{12}\,\text{M}_\odot/h$ in $M_{200\text{c}}$, is $0.944 \pm 0.017$ and $0.732^{+0.025}_{-0.020}$. Counting satellites as well as centrals, the model falls short of the former ($0.84$ in the stock and calibrated configurations alike), since satellites are less often quiescent than centrals, which alone give $0.95$, in agreement with the measurement. Above $10^{12}\,\text{M}_\odot/h$, the model exceeds the measurement ($0.80$ and $0.81$), quenching galaxies of haloes between $10^{12}$ and $10^{13}\,\text{M}_\odot/h$ more completely than is observed. Both offsets are common to every configuration of the sweep, and so are inherited from the base model rather than caused by \textsc{Aurora}.

At $3.4 \le z < 4.2$, \citet{Straatman2014} find $34 \pm 13$ per cent of galaxies above $10^{10.6}\,\text{M}_\odot$ in \textsc{Zfourge} to be quiescent, and at $3 < z < 4$, \citet{Muzzin2013} find 16 per cent of \textsc{UltraVISTA} galaxies above their completeness limit of $10^{10.94}\,\text{M}_\odot$ to be quenched. With galaxies selected on stellar mass alone, as in both surveys, and pooled over the interval of \citet{Straatman2014}, the quiescent fraction is $8.3$ per cent in the stock model and $8.5$ per cent in the calibrated configuration (\cref{fig:fqhalo}). This shortfall by a factor of two to four is inherited from the base model, and \textsc{Aurora} leaves it unchanged.

\begin{figure}
\includegraphics[width=\columnwidth]{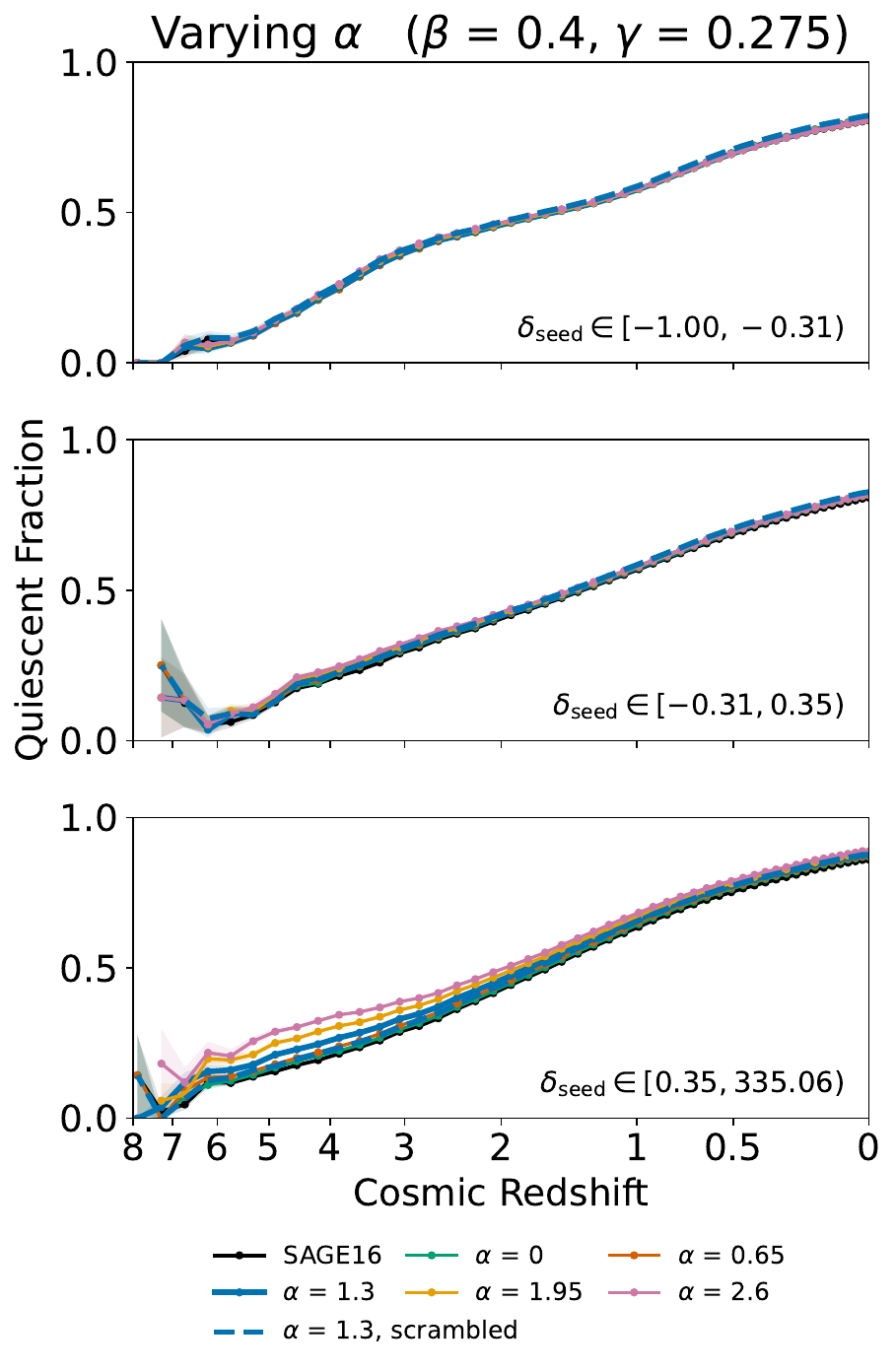}
\caption{Quiescent fraction of seeded galaxies in host haloes above $10^{12}\,\text{M}_\odot/h$, split into terciles of the overdensity at which each galaxy's own black hole was seeded. From top to bottom, each panel represents the sparse, average and dense environments of \Cref{sec:evolution}. $\check{\alpha}$ is varied, with the stock \textsc{Sage} model in black and the calibrated configuration rerun with scrambled seeding environments dashed. The response is concentrated in the densest tercile, while the sparser terciles lie close to the stock model and move little with the coupling: because $\ln(1+\delta)$ is centred on zero, the overdensity coupling pivots the seed spectrum between environments, and the quiescence it adds is largest in the densest of them. With the environments scrambled, the curves lie close to the stock model in every tercile at $z \gtrsim 2$, and above it by no more than $0.05$ towards $z = 0$.}
\label{fig:fqenv_alpha}
\end{figure}

\begin{figure}
\includegraphics[width=\columnwidth]{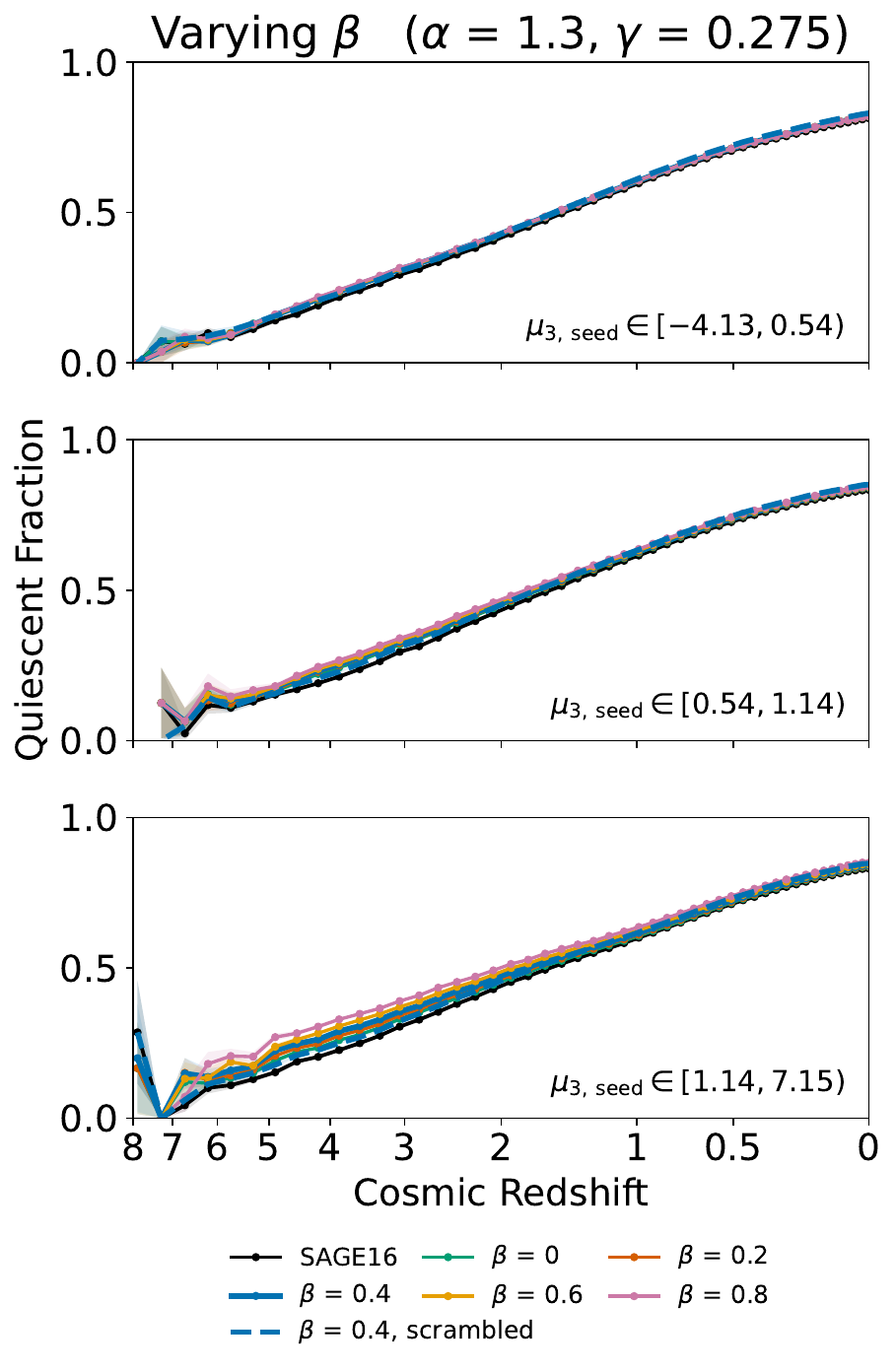}
\caption{As \cref{fig:fqenv_alpha}, with $\check{\beta}$ varied and the galaxies split into terciles of the radial skewness of their seeding environment, the quantity to which that index couples. Among the galaxies of \cref{tab:metrics}, the enhancement grows with skew, steepening the gradient across the terciles from $\Delta\Theta = +0.072$ at the calibrated value to $+0.16$; unlike the overdensity coupling, however, it also steepens the gradient across \emph{overdensity} terciles, by more than half as much (\cref{sec:quiescence}).}
\label{fig:fqenv_beta}
\end{figure}

\begin{figure}
\includegraphics[width=\columnwidth]{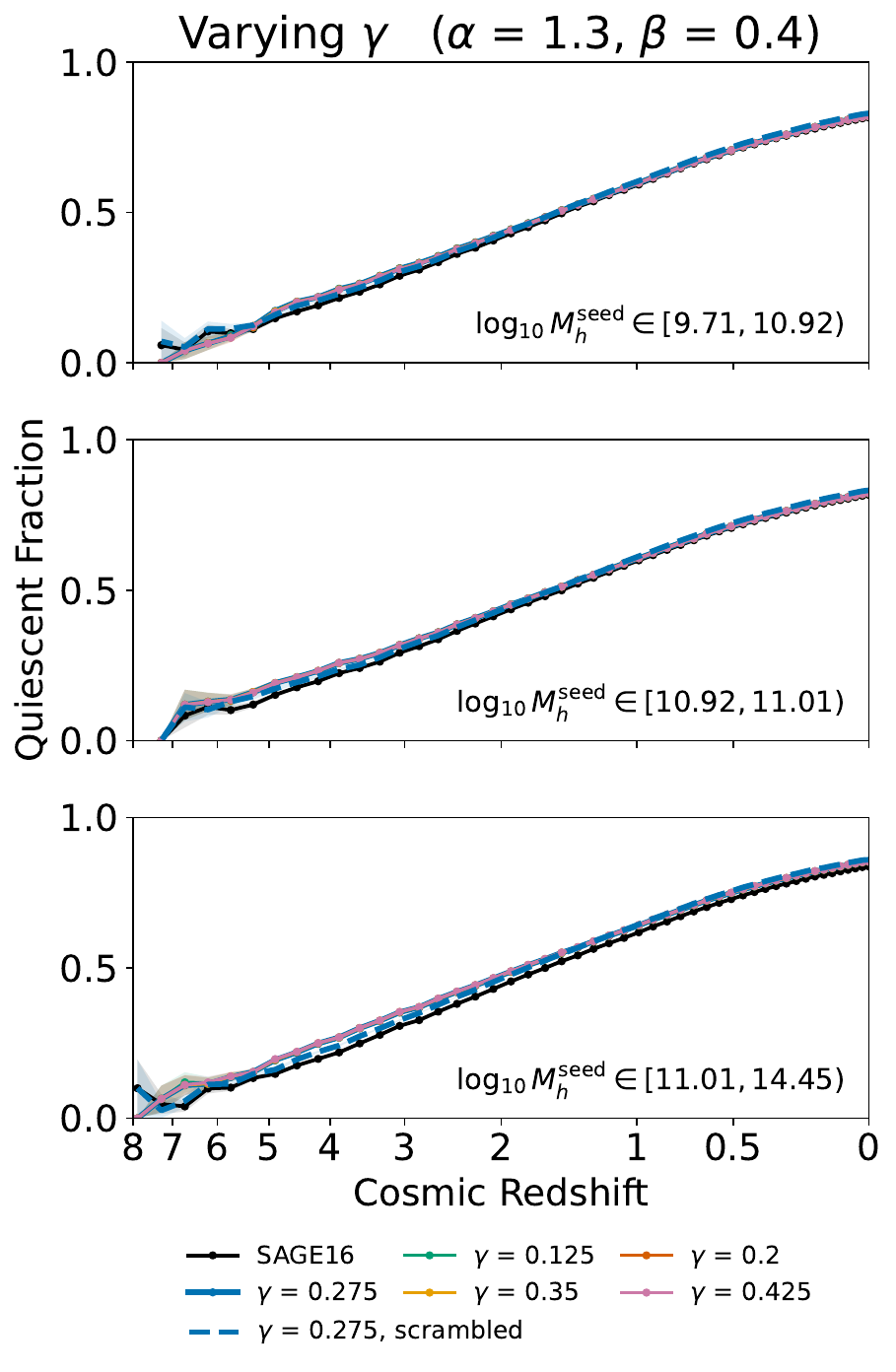}
\caption{As \cref{fig:fqenv_alpha}, with $\check{\gamma}$ varied and the galaxies split into terciles of their seeding halo mass. The curves are indistinguishable from the calibrated configuration in every tercile, as seeds are assigned at the moment of eligibility, so the seeding halo mass has almost no spread across the bulk of the population; the coupling shows instead on the small tail of haloes which cross the threshold with substantial overshoot (\Cref{sec:sweeps}), which a tercile split of the whole population cannot isolate.}
\label{fig:fqenv_gamma}
\end{figure}

The manner of the enhancement differs between the two active couplings, and this is where the distinction between cosmic variance and interactions becomes observable rather than merely formal. \Crefrange{fig:fqenv_alpha}{fig:fqenv_gamma} split the same galaxies into terciles of the seeding epoch quantity to which the varied index couples (overdensity for $\check{\alpha}$, radial skewness for $\check{\beta}$ and seeding halo mass for $\check{\gamma}$), with each galaxy taking the environment in which its own black hole was seeded. Each coupling steepens the gradient in its own variable. At $z \simeq 3.1$, raising $\check{\alpha}$ from the calibrated $1.3$ to $2.6$ moves the two sparser $\delta$ terciles by only $+0.013$ and $+0.021$, to $\Theta_1 = -0.015$ and $\Theta_2 = +0.009$, while raising the densest from $\Theta_3 = +0.10$ to $+0.34$. The environmental gradient across the $\delta$ terciles steepens accordingly, from $\Delta\Theta = +0.13$ to $+0.36$. Raising $\check{\beta}$ from $0.4$ to $0.8$ leaves the least asymmetric $\mu_3$ tercile almost unchanged, at $\Theta_1 = +0.051$ against $+0.042$, and raises the most asymmetric from $\Theta_3 = +0.11$ to $+0.22$, steepening the gradient across $\mu_3$ terciles from $\Delta\Theta = +0.072$ to $+0.16$. Varying $\check{\gamma}$ reproduces the calibrated values in every tercile to within $0.007$ (\cref{fig:fqenv_gamma}).

The randomisation test isolates what that gradient owes to environment rather than to structure formation. With the environments scrambled, the gradient across $\delta$ terciles falls from $\Delta\Theta = +0.13$ to $-0.008$: none of it survives the loss of the correspondence between a halo and its surroundings. The sign structure goes with it; the calibrated model places a deficit of quiescent galaxies in the sparsest tercile and an excess in the densest, $\Theta_1 = -0.028$ against $\Theta_3 = +0.10$. The scrambled run returns $+0.009$ and $+0.001$, within $0.01$ of the stock model in every overdensity tercile. The gradient across $\mu_3$ terciles falls likewise, from $\Delta\Theta = +0.072$ to $-0.008$.

The two couplings are not, however, equally selective. Measured in terciles of $\mu_3$, raising $\check{\alpha}$ to $2.6$ lifts all three by $+0.19$, $+0.21$ and $+0.19$ relative to the calibrated configuration, leaving the skew gradient unchanged at $\Delta\Theta = +0.072$. Measured in terciles of $\delta$, raising $\check{\beta}$ to $0.8$ lifts all three by $+0.009$, $+0.035$ and $+0.060$, steepening the overdensity gradient from $\Delta\Theta = +0.13$ to $+0.18$. The overdensity coupling is therefore neutral in skew, while the skew coupling carries a secondary overdensity dependence.

Since $\ln(1+\delta)$ is centred on zero, the overdensity coupling steepens the environmental gradient, adding quiescence in nodes far more than in voids, and does little to a population selected on skew beyond a near-uniform shift. As $\mu_3$ is positive for nearly three quarters of seeding events, the skew coupling instead raises seed masses almost everywhere, yet the quiescence it adds still grows with overdensity: it barely lifts the sparsest $\delta$ tercile, by $+0.009$ against $+0.013$ for $\check{\alpha}$, while discriminating most strongly among the asymmetric environments it is designed to select. The two should therefore be distinguishable by future observations: one predicts a steepening contrast between voids and nodes, while the other predicts an excess of early quiescent systems in interaction-rich neighbourhoods, which also grows with the ambient density.

At $z \simeq 3.1$ the stock model's quiescent fraction falls from $0.44$ in the sparsest $\delta$ tercile to $0.39$ in the densest, reflecting the environmental dependence already present in the base model. The calibrated configuration moves both ends, and in opposite directions: it raises the densest tercile by $+0.10$, to $0.50$, and lowers the sparsest by $0.028$, to $0.42$. Together, they make an environmental gradient of $\Delta\Theta = +0.13$, more than enough to reverse the stock model's decline in quiescent fraction from the sparsest tercile to the densest. Measured instead against the scrambled control, which shares the seed mass spectrum, the two shifts are $-0.038$ and $+0.10$, illustrating that both are environmental. Only the overdensity term can produce shifts of opposite sign.

Because $\ln(1+\delta)$ is centred on zero, it lowers as many seeds as it raises (\Cref{sec:couplings}), whereas the skew term is positive for most seeding events, and the halo mass term is small for most of them. Accordingly, in the calibrated configuration, every tercile of $\mu_3$ and of seeding halo mass is raised (\cref{tab:metrics}), and only the overdensity split changes sign. An excess of early quiescent systems in overdense environments, accompanied by a deficit in voids, is therefore the key prediction of environmental seeding: a joint signature that no uniform change in the efficiency of feedback can reproduce. The gradient which the calibrated configuration adds rises from $\Delta\Theta = +0.13$ at $z \simeq 3.1$ to $+0.18$ at $z \simeq 4.2$ (\Cref{sec:disc_signal}), as self-regulation has had less time to erase the seed dependence of the feedback, and it justifies a direct observational test of environmental seeding.

\section{Discussion}
\label{sec:discussion}

\subsection{A persistent and verifiable seeding signal}
\label{sec:disc_signal}

The $N$-body simulation used in this work demonstrates \textsc{Aurora}'s key capabilities and the sensitivity of its parameters, and the scientific results of the \textsc{Aurora} programme will come from its application to high resolution, large volume simulations. What this paper establishes is the property that makes such applications worthwhile: an environmentally graded seeding prescription leaves a real, measurable signal in the galaxy population long after the seeding events themselves, which may be used in future to constrain black hole seeding mechanisms from observations. Self-regulated accretion drives individual black hole masses towards an attractor set by the host, yet the population level budget of \cref{tab:metrics} shows the seeds persisting in bulk, because what one lineage loses to accretion the population recovers through the coalescence of seeds along its merger tree (\Cref{sec:synthesis}). Furthermore, the near-perfect degeneracy of the $z=0$ stellar-halo mass relation and median galaxy-black hole mass relation across the sweep (\Cref{sec:sweeps}), which the black hole mass function does not share (\cref{fig:sweep_bhmf}), does not extend to earlier epochs. The quiescent fraction of seeded galaxies in massive haloes carries an enhancement which grows with the overdensity and skew couplings, reaching $\Theta = +0.28$ at $\check{\alpha} = 2.6$ and $z \simeq 3.1$, and an environmental gradient across overdensity terciles which grows from $\Delta\Theta = +0.13$ in the calibrated configuration to $+0.36$ at $\check{\alpha} = 2.6$ (\Cref{sec:quiescence}).

The same argument has been made of the black hole mass function itself. \citet{Taylor2025} measure it for broad-line \acp{agn} over $3.5<z<6$ and find a featureless power law, reading it as a seeding signature already lost by $z \simeq 5-6$. \citet{Jeon} account for that loss: light and heavy seeds grown forward in the same merger trees reproduce the function alike, with a spread in accretion rate erasing the memory of the seed mass scale and the two separating only above $z \simeq 10$. That statement is about a formation channel mass at an epoch this simulation does not reach. \textsc{Aurora}'s seeds enter at the eligibility threshold inside the interval they measure, and what is graded by environment is that effective seed. Where the observables meet, the two agree, with \textsc{Aurora}'s mass function departing from a single power law over their interval by 0.25 dex rms against 0.35 dex for the stock model; the comparison is of shape alone, since theirs counts only broad-line \acp{agn}. Yet with the mass functions similar, the calibrated configuration carries an environmental gradient across overdensity terciles of $\Delta\Theta = +0.13$ at $z \simeq 3.1$ and $+0.18$ at $z \simeq 4.2$ (\Cref{sec:quiescence}); the galaxies in which the former is measured were seeded a median 1.1 Gyr earlier, an interval far longer than the mixing time of their estimate.

This information is absent from how many black holes of each mass there are, and present in which galaxies they quench, and in which surroundings. Scrambling the environments leaves the seed mass spectrum and the black hole mass function intact, with the total black hole mass density agreeing with the calibrated run to $0.3$ per cent, but removes the environmental gradient in quiescence entirely, together with the quiescent enhancement at $z \simeq 3.1$ (\cref{tab:metrics}). The mass function is insensitive to the control because it counts black holes by mass and discards which halo, in which surroundings, each one occupies: it marginalises over precisely the information the prescription encodes. What the calibration constrains is therefore the seed mass distribution at a given halo mass and epoch; what it does not constrain, and what the gradient in quiescence measures, is which halo receives which seed.

Divided by environment, the mass function recovers part of that information. In terciles of the overdensity at $z = 0$,\footnote{The overdensity at $z = 0$ is recomputed more crudely than \textsc{Aurora}'s own, with the same Gaussian kernel of $1\,\text{cMpc}/h$ evaluated on a $512^3$ mesh and not truncated. The split by present-day overdensity nonetheless concentrates the excess in the densest tercile, as the split by seeding overdensity does for quiescence (\Cref{sec:quiescence}).} the calibrated configuration holds $6.4$, $5.5$ and $17$ per cent more black holes above $10^{7}\,\text{M}_\odot/h$ than the stock model, from the sparsest tercile to the densest, and $27$ per cent more black hole mass in the densest, against $2.7$ per cent less in the sparsest. The mass function of black holes by environment at $z = 0$ is therefore a prediction in its own right.

The fact that the discriminating information should lie in configuration space rather than in the global mass function is a conclusion others are reaching independently. \citet{Wang2026} find the candidate origins of the compact red population to separate in their large scale clustering, with the direct collapse route predicting the strongest bias and a primordial one predicting none above that of the matter field, whereas their mass functions do not separate at all. \citet{Shen2026} reach the same point from the other direction. In \textsc{Lumina} \citep{Zier2026}, a $(500\,\text{cMpc})^3$ radiation-hydrodynamical simulation whose seeding depends on halo mass alone, the cross-correlation of \acp{lrd} with star forming galaxies is almost identical to the galaxy auto-correlation, and matches the observed amplitude for host haloes slightly above $10^{11}\,\text{M}_\odot$. The authors caution that a formation channel tied to environment or assembly history could change that clustering at fixed halo mass. Environmental seeding is one such channel, and \textsc{Aurora} adds it to the halo mass threshold on which the seeding of \textsc{Lumina} rests.

The persistence of the environmental gradient is of direct value to efforts to infer the black hole seeding history from observations, whether through the demography of high redshift \acp{agn}, whose abundance \textit{JWST} spectroscopy now places well above extrapolations of the quasar luminosity function \citep{Inayoshi2020, Harikane2023}, the gravitational wave background of merging seeds \citep{Liu&Inayoshi}, or the coevolution of black holes and their hosts, whose evolving mass ratio is itself a diagnostic of the seeding pathway \citep{Habouzit, Ding2023, Pacucci, Mezcua, Scoggins2024}. The two environmental couplings imprint separable signatures: a steepening contrast between voids and nodes for the overdensity coupling, and a one-sided enhancement tracking local asymmetry for the skew coupling (\cref{fig:fqenv_alpha,fig:fqenv_beta}).

The observational material for such tests is accumulating. Spectroscopically confirmed quiescent galaxies are now found within protoclusters and dense structures at $z \gtrsim 3.4$ \citep{Tanaka2023, Ito}, representing precisely the overdense, asymmetric configurations in which the couplings measured here are strongest, and the clustering of the earliest galaxies is itself now being measured, over the contiguous areas needed to separate it from cosmic variance \citep{Weibel3}. Hydrodynamical models point the same way: in \textsc{Colibre}, the black hole masses and star formation efficiencies which distinguish \acp{mqg} from star forming galaxies of equal mass are traced to their environments before quenching, on local ($0.3$ cMpc) to intermediate ($1.0$ cMpc) scales, where overdense regions drive enhanced inflows and hence more powerful feedback \citep{ChandroGomez2}.

\begin{figure}
\includegraphics[width=\columnwidth]{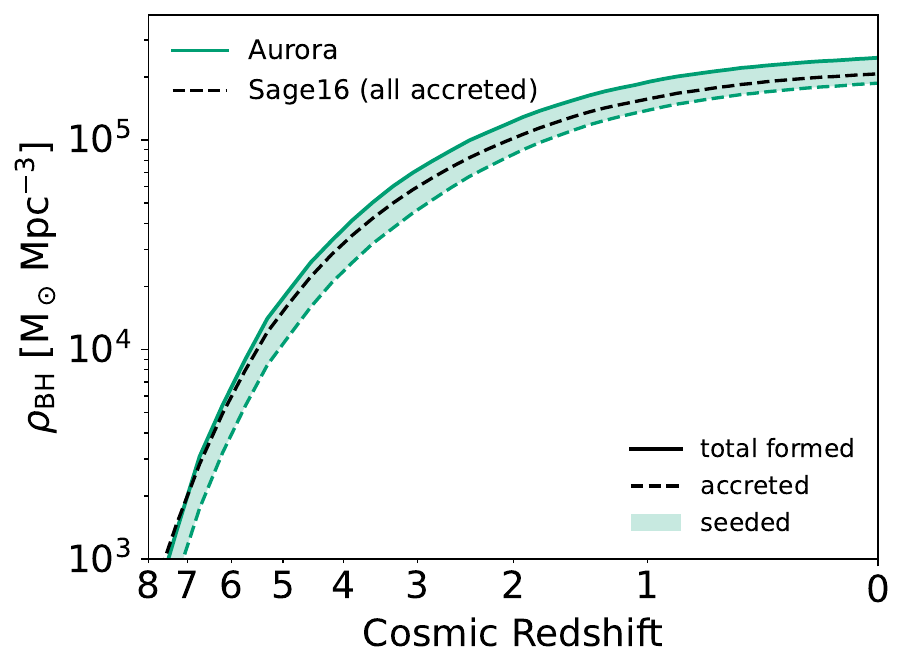}
\caption{The So\l{}tan decomposition of the \textsc{Aurora} black hole mass budget, after \citet{Volonteri2010}, shown as cumulative black hole mass density against redshift. The solid curve represents all mass ever formed, seed together with accreted; the dashed curve is the accreted component alone, which is the quantity which So\l{}tan-type arguments constrain, and the shaded band between them is the seed contribution. The stock \textsc{Sage16} model is drawn dashed in black: since it carries no explicit seed, its total and accreted densities are the same curve, and it is therefore directly comparable to the dashed \textsc{Aurora} curve, above which it lies at every epoch. \textsc{Aurora} accumulates more black hole mass in total while accreting less of it.}
\label{fig:soltan}
\end{figure}

Seeding explicitly, where the base model does not, raises the question of whether \textsc{Aurora} places black hole mass outside the accretion channel constrained by the argument of \citet{Soltan}, which equates the black hole mass density accumulated through accretion with the integrated luminosity density of quasars, for an assumed radiative efficiency. This is not the case; gating growth on the presence of a seed suppresses the bootstrapping which the stock channel performs below the eligibility floor, and \textsc{Aurora} injects less quasar mode accretion than \textsc{Sage16} on identical trees: the seeds substitute for accretion rather than adding to it. In any case, the \citet{Soltan} argument constrains the accreted mass density rather than the total mass locked into black holes; \citet{Volonteri2010} draws this distinction explicitly and represents it by plotting the two as separate curves. The seed budget fraction $\eta$ manifests as the separation between them, and \Cref{fig:soltan} shows that decomposition for \textsc{Aurora}: the accreted curve lies below the stock model throughout, while the total, which carries the seed contribution, exceeds it. The separation is widest at early times, with $\eta$ peaking near $0.46$ at $z \simeq 7$ before declining to $0.29$ by the present day as accretion accumulates.

\subsection{Extending the prescription}
\label{sec:disc_extend}

The overdensity coupling of \Cref{eq:BHseeding} is not merely an empirical convenience, and two recent radiation hydrodynamic calculations identify the physical quantities for which it stands in. \citet{Chon2} find that the Lyman-Werner field which permits a heavy seed to form is highly inhomogeneous: at the epoch of the first heavy seed in their volume the mass weighted intensity has reached $J_{21} \simeq 10^{3}$, while the volume weighted average remains of order unity, with the difference arising because strong irradiation is confined to dense, clustered star forming structures. The quantity which fixes the seed is therefore itself a steep function of local overdensity. \citet{Prole2026} approach the same point from a different direction, making heavy seed formation conditional on the mass inflow rate reaching a halo's centre, so that a heavy seed appears only where the central supply is high, and not where it never becomes so.

Our own catalogue reproduces that association. At fixed seeding halo mass and fixed seeding redshift, central galaxies seeded in the densest tercile of $\delta$ have since assembled their haloes faster than those seeded in the sparsest, by factors of $1.1$ to $3.1$, and the separation widens towards high redshift, reaching $3.1$ for seeds planted at $z = 7-10$. The overdensity term is thus a proxy, available from the dark matter alone, for the rapid assembly those calculations identify as the condition for heavy seed formation. The same dependence survives into larger volumes: \citet{Petkova2026} regulate seeding by the ionising feedback of the sources already formed, so that where a halo sits relative to its neighbours decides whether it seeds at all. The correspondence is qualitative rather than quantitative: \citet{Prole2026} evaluate an inflow rate within $10$ parsecs, four orders of magnitude below the scale on which environment is measured here, and we do not claim that $\check{\alpha}$ measures it.

\citet{Jeon2026} offer the accretion-side counterpart to this account. Arguing from halo abundance alone, they find that cold-mode inflow is required to supply the black hole mass observed at high redshift, and that overmassive systems could arise where that inflow feeds the black hole rather than forming stars. As an explanation of overmassive black holes this competes with environmental seeding, as does the tidal stripping of satellite hosts, which \citet{IzquierdoVillalba2026} find raises the mass ratio at low redshift, in a semi-analytic model on the Millennium trees, by removing stellar mass while leaving the black hole untouched. In the light of \Cref{sec:pathway}, where seeding moves the high ratio tail towards the observed locus without reaching it, the three are better read as complementary, acting respectively on the seed, on its subsequent accretion, and on the stellar mass against which it is measured. The rapid assembly which the overdensity at seeding predicts is the supply of such inflow, so the environments in which \textsc{Aurora} plants heavier seeds are those in which that growth should then be sustained. \textsc{Aurora} does not test the accretion side of this picture, and we make no such claim: \textsc{Sage} cools gas spherically, with no representation of the anisotropic streams through which cold-mode accretion proceeds. What the correspondence suggests instead is the natural next step for the prescription, coupling the seed to the rate at which its halo is assembling rather than to its instantaneous overdensity, which the architecture described below accommodates directly.

\Cref{eq:BHseeding} is one member of a family. The architecture which evaluates it, a per-snapshot environment cache attached to the merger trees read by a gated seeding module, accepts any quantity computable from the $N$-body data, and equally any quantity derived within the \ac{sam} itself. In future work, these additional quantities may be used to create a more sophisticated black hole seeding model.

One physically pressing extension is metallicity. Population III remnants form from metal-poor gas \citep{Bromm2003}, while the massive, stable seeds of direct collapse require the deep potentials assembled in rare, rapidly growing haloes \citep{Inayoshi2020, Hopkins, Regan2024}. A prescription in which the seed mass, or the choice among seeding mechanisms, responds jointly to the gas phase metallicity carried by the \ac{sam} and to the halo growth history, would promote \textsc{Aurora} from a single environmental grading to a genuine multi-channel seeding model, in the spirit of the channel-resolved prescriptions now implemented semi-analytically and in cosmological hydrodynamics \citep{Spinoso2023, Bhowmick2024, Bhowmick2026}. The dominant channels would vary across environment exactly as the physics dictates, and, unlike those implementations, would remain applicable to gravity-only volumes. A condition of this kind would also close a window which \textsc{Aurora} shares with threshold seeding generally, in which any halo crossing the eligibility mass is eligible for a seed, at any epoch. \citet{Shen2026} identify the same feature of \textsc{Lumina} as a possible cause of the continued rise of its \ac{lrd} abundance below $z \simeq 4$, where the observed abundance declines.

The halo mass coupling deserves separate comment, because it is the one term of \Cref{eq:BHseeding} whose effect is confined to the high mass tail. Seeds are assigned at the moment a lineage first crosses $\check{N}_h^\text{min}$, so the ratio entering the third term is close to unity for almost every event. The term is only significant for haloes which cross the threshold with substantial overshoot, and at $\check{N}_h^\text{min} m_p \simeq 10^{11}\,\text{M}_\odot/h$ such haloes are rare.

The point generalises beyond the coupling itself. Since \Cref{eq:BHseeding} is specific to the eligibility mass, the particle mass sets the mass scale of the entire spectrum, and the offset which $\check{\gamma}$ contributes in this volume is proportional to it. Read physically, that offset stands in for the growth accumulated over the interval between the formation of a seed and the snapshot at which its halo first becomes resolved. This interval is long at $\check{N}_h^\text{min} m_p \simeq 10^{11}\,\text{M}_\odot/h$, and shrinks as the resolution improves. A prescription calibrated at one particle mass therefore cannot simply be transplanted to another; the normalisation $\check{M}^0_\bullet$ must be recalibrated with it, and the physical interpretation of $M_\bullet$ shifts as the offset closes.

At \textsc{Millennium} resolution, the threshold of $\check{N}_h^\text{min} = 115$ particles clears the convergence limits established for merger trees of this kind: \citet{Poole} find halo abundances converged to $\sim\!5$ per cent for $n_p \gtrsim 75$, and substructure to $\sim\!10$ per cent for $n_p > 100$, at mass resolutions $m_p \lesssim 10^{9}\,h^{-1}\text{M}_\odot$. Their stricter threshold, $n_p \gtrsim 300$, above which haloes cease to lose their descendants, does not bear on seeding: $\delta$ and $\mu_3$ are evaluated in the snapshot at which a halo first becomes eligible, independently of its merger history (\Cref{sec:aurora_cache}), and a lineage which later strays forfeits the same growth in the calibrated and base runs, which share their trees. The differential statistics reported here are therefore robust to straying haloes, even where absolute abundances are not.

Two things change at higher mass resolution. The eligibility mass falls with the particle mass, so a halo may grow by a large factor between snapshots while crossing it, and the distribution of $M_h^\text{seed}$ acquires real width. In such a regime, the term ceases to be a weak scaling with halo mass and becomes a measure of how rapidly the halo was assembling when its black hole formed. Additionally, high-resolution data will reliably provide the depth of the potential well at seeding, which distinguishes seeding channels. The halo mass term is therefore the one through which a channel dependence would most naturally enter, and the one whose calibration requires simulations which resolve the progenitors of the seeding haloes; either high resolution $N$-body volumes \citep{Uchuu}, or the sub-kiloparsec dwarf simulations which resolve the conditions under which each channel operates \citep{Rey2025}.

Raising $\check{\gamma}$ to the top of its explored range leaves $f_q$, $\Theta$ and every $\Theta_i$ of \cref{tab:metrics} within $0.007$ of the calibrated configuration, while raising $\check{\alpha}$ to the top of its own moves the quiescent fraction of seeded galaxies in massive haloes by $0.20$. The halo mass term is thus measurable in this volume but weak in its galaxy population, and it is weak for the reason set out above: the distribution it multiplies has almost no width. That is a measurement of how little dynamic range the term is given at this resolution, not of how much the coupling is worth. What it is worth is a question the same prescription will answer when applied where the range exists.

\subsection{The high redshift frontier}
\label{sec:disc_frontier}

The strongest motivation for \textsc{Aurora} lies at high redshift, where the objects of interest are rare, extreme, and systematically absent from conventional models. \Acp{mqg} are now established observationally to $z \simeq 5$ and beyond \citep{Carnallnature, Valentino2023, deGraaff, Nanayakkara, Weibel, Ji2026, Zhang2026}, in numbers which most contemporary simulations underpredict by an order of magnitude or more, with quenching mechanisms at these epochs only beginning to be characterised in the models that produce them at all \citep{Xie, Kimmig, RemusKimmig, Seeyave2025, Szpila, ChandroGomez1, ChandroGomez2, Chon}. The \acp{lrd} present a sharper puzzle still: abundant, compact, apparently hosting overmassive black holes \citep{Greene, Kokorev2024, Kocevski2025, Labbe, Matthee2025, Taylor2025, Kapoor2026}, yet with an interpretation which remains unresolved \citep{PerezGonzalez2024, Inayoshi2025b, deGraaff2026, Inayoshi2025, JiLRD, Weibel2}, including whether their black holes are as massive as virial estimates suggest \citep{Rusakov2026, Brazzini2025}. Furthermore, the luminous $z \gtrsim 6$ quasars \citep{Fan2023, EuclidCollaboration2026} and the extreme early assembly of systems such as the \ac{mqg} of \citet{Glazebrook2024}, whose \ac{sfh} implies a Milky Way stellar mass in place by $z \simeq 11$, bracket the same problem from above: some environments built massive black holes and massive galaxies far faster than median-calibrated prescriptions allow. The two are now seen together in single objects: among twelve moderate-luminosity quasars at $z > 6$, the two host galaxies carrying post-starburst features match the sizes and stellar mass surface densities of quiescent galaxies at $z \simeq 4-5$, placing the compaction and quenching route already in the reionisation era \citep{Ding2025}.

These are, without exception, tail phenomena, and it is in the tails where \textsc{Aurora} differs most from what it replaces. A fixed seed mass, or one tied to halo mass alone, assigns the rarest environments unremarkable black holes; the environmental grading instead concentrates seed mass exactly where $\delta$ and $\mu_3$ are extreme, producing the seed-dominated pathway and the extended high ratio tail of the galaxy-black hole mass relation (\Cref{sec:pathway}). As the prescription operates on dark matter information alone, it can populate gigaparsec volumes at negligible cost, permitting a class of test unavailable to resolved simulations. Future work may invoke forward-modelling the abundance, host masses, environments, and clustering of overmassive black hole systems under an \ac{agn} interpretation of the \acp{lrd}, in the spirit of the \textsc{Astrid} to \textsc{Brahma} comparison of \citet{LaChance2025} but across volumes a hydrodynamical suite cannot reach. It remains to be seen whether any coupling within the explored family reproduces the observed census, whose inferred bolometric luminosity function exceeds the ultraviolet-selected quasar luminosity function a hundredfold \citep{Akins2025}.

The empirical mapping of \citet{Shen2026} is built for such a test. Designed to transfer to semi-analytic models, it converts black hole accretion into multi-band luminosity functions, and treats a fixed fraction of the black holes still within a factor of ten of their seed mass as \acp{lrd}, reproducing both the observed \ac{lrd} luminosity functions and their clustering in \textsc{Lumina}. Read relative to each black hole's own seed, that criterion is a condition on seed retention, $\zeta \geq 0.1$, the complement of the accretion-dominated regime of \Cref{sec:pathway}, such that under \textsc{Aurora} the incidence of the \ac{lrd} phase would inherit the environmental grading of the seeds. A systematic failure would itself be informative, disfavouring black hole dominated interpretations at the population level; success would yield the seeding parameters as measurements.

The abundance of \acp{mqg} is the clearest such test, and one which this volume cannot carry out, since the test requires two things which \textsc{Millennium} does not resolve. The first is a highly resolved formation history for each host, followed from an early seed, so that black hole and host grow together from the epoch at which the host's stars form, under a control of their coevolution which only resolved progenitors and finely sampled merger trees provide. The second is the primordial environment of the host's progenitors, which only haloes far below the \textsc{Millennium} resolution limit can trace at such early epochs. In \textsc{Millennium}, a halo becomes eligible only at $\simeq 10^{11}\,\text{M}_\odot/h$, at a median redshift of 4.5 for the main progenitors of the massive centrals of $z = 0$, and it receives in a single step a seed which stands in for the growth its progenitors accumulated unresolved. The feedback which that black hole powers then acts on a host which has yet to build its stellar mass, and the centrals which \textsc{Aurora} adds to the quiescent population quench prematurely. This early quenching is a consequence of the resolution, not of the seeding law. The number density of \acp{mqg} is therefore the principal target of the next application of \textsc{Aurora}, at a resolution which addresses both requirements, and not of this work.

Our immediate application of \textsc{Aurora} will be to \textsc{Shin-Uchuu} \citep{Uchuu}, whose particle mass of $8.97 \times 10^{5}\,\text{M}_\odot/h$ is nearly a thousand times smaller than that of \textsc{Millennium}. Its resolution places the eligibility threshold at $\sim\!10^{8}\,\text{M}_\odot/h$ rather than $\sim\!10^{11}$, such that seeding begins earlier, in the progenitors of the haloes seeded here, and in environments traced by the low mass haloes which populate the primordial field. With the additional benefit of a highly resolved merger tree time domain, we will evaluate the key growth mechanisms which operate in different environments at high redshift, setting the rapid, seed-retaining channel against the accretion-dominated one, and follow the evolutionary pathways of the earliest massive black hole systems, \acp{mqg} and \ac{lrd} candidates, through the reionisation epoch, and down to $z = 0$. At \textsc{Millennium} resolution, the majority of seeding events occur after reionisation, reflecting a limitation of mass resolution that volume cannot address. Seeding at those redshifts, and resolving the growth which follows, is the modelling counterpart to the observational programmes this frontier defines: the search for a formation channel signature above $z \simeq 10$ \citep{Taylor2025}, the census of \acp{lrd} \citep{Akins2025}, and the earliest quiescent systems \citep{Glazebrook2024}. Extending the same prescription to the $2\,\text{cGpc}/h$ \textsc{Uchuu} volume would then place the rarest systems, which no smaller box contains, on a statistical footing.

\section{Conclusions}
\label{sec:conc}

In this work we have presented \textsc{Aurora}, a black hole seeding module for the \textsc{Mimic} semi-analytic framework, in which the seed mass assigned to a newly eligible halo is graded by the environment in which that halo forms. The motivation is the persistent failure of conventional models to reproduce the abundance of \acp{mqg} and overmassive black holes now observed at high redshift \citep{Carnallnature, Valentino2023, deGraaff, Maiolino, Nanayakkara, Labbe, Weibel, Ji2026, Zhang2026}, together with the finding, in \textsc{Thesan} \citep{Chittenden3} and \textsc{Colibre} \citep{ChandroGomez2} alike, that the earliest such systems assemble in dense environments and grow their black holes early and rapidly. This work tests the environmental side of that picture, asking whether environmental seeding leaves a measurable imprint on the galaxy population; the abundance of \acp{mqg} is left to an application at higher resolution (\Cref{sec:disc_frontier}). Rather than assigning every halo a fixed seed, \textsc{Aurora} couples the seed mass in log space to the local overdensity $\delta$, the mass-weighted radial skewness $\mu_3$ of the surrounding halo distribution introduced by \citet{Chittenden}, and the halo mass at seeding, through the three indices of \Cref{eq:BHseeding} (\Cref{sec:aurora_seedmass}). The environmental metrics are computed once per snapshot and cached (\Cref{sec:aurora_metrics,sec:aurora_cache}), so the prescription requires only the dark matter component of the simulation, and can be applied to volumes far beyond the reach of hydrodynamics.

We calibrated the prescription against the \textsc{Sage16} \acl{sam} \citep{Croton2006, Croton2016} run on the \textsc{Millennium} merger trees, matching the $z=0$ black hole mass function and the count of black hole hosts (\Cref{sec:calibration}), and explored the behaviour of each coupling through one-dimensional sweeps about the calibrated configuration (\Cref{sec:sweeps}). We then followed the consequences forward in time, from the seed mass spectrum to the assembly of massive black holes and the quiescent galaxy population (\Cref{sec:evolution}), using the metrics defined in \cref{tab:metrics}.

Our findings can be summarised as follows:

\begin{itemize}
\item The three couplings act in structurally different ways, and the difference follows from the distributions of the quantities which they multiply, rather than from the couplings themselves (\Cref{sec:couplings}). Because $\delta$ is normalised to the geometric mean density, $\ln(1+\delta)$ has a median of $+0.002$ across seeding events and the overdensity coupling pivots the seed spectrum, boosting seeds in nodes and suppressing them in voids. The skewness $\mu_3$ is positive for 72 per cent of events, so the skew coupling instead raises seed masses almost everywhere. The two are nearly orthogonal, with a Spearman rank correlation of $-0.038$.

\item Massive black holes in \textsc{Aurora} are synthesised from many environmentally graded seeds rather than grown from one (\Cref{sec:synthesis}). Seeds supply a fraction $\eta = 0.29$ of all black hole mass formed in the calibrated configuration, rising to $0.74$ at $\check{\alpha} = 2.6$, where seeding rather than accretion supplies the majority of the budget. Individual seed retention $\zeta$ falls from unity in haloes below $10^{11}\,\text{M}_\odot/h$ to $0.018$ above $10^{12}\,\text{M}_\odot/h$ at $z=0$, but rises again towards high redshift, reaching $0.10$ at $z \simeq 4.2$. Environmental seeding therefore remains legible in the population long after it has ceased to be legible in any individual black hole.

\item A seed-dominated growth pathway emerges, distinct from the accretion-dominated majority (\Cref{sec:pathway}). A minority of massive centrals retain more than half of their black hole mass as their original seed; these were seeded in markedly denser and more asymmetric environments than the accretion-dominated majority ($\delta = +1.6$ against $+0.06$ at $z \simeq 2.1$, and $+2.3$ against $+0.19$ at $z \simeq 3.1$), and respond to the couplings by up to $+0.55$ dex in black hole mass where the majority do not shift noticeably. This channel corresponds closely to the dense environment assembly identified for the earliest \acp{mqg} in the \textsc{Thesan} weak feedback experiment \citep{Chittenden3}, weak because of an implementation error in the public release \citep{Garaldi2026}, and to the earlier seeding and faster black hole growth which distinguish them in \textsc{Colibre} \citep{ChandroGomez2}, recovered here from dark matter information alone. That a gravity-only model reproduces a channel found in two hydrodynamical simulations under contrasting feedback regimes is consistent with the conclusion of \citet{Chittenden3} that its conditions for black hole synthesis are set by assembly history rather than by the subgrid physics, even where the black hole growth those conditions enable is not.

\item \textsc{Aurora} produces some black holes overmassive relative to their hosts, in a configuration which qualitatively resembles that inferred for compact \acp{agn} and \acp{lrd} at $z \gtrsim 4$ \citep{Harikane2023, Maiolino, Ubler, Kocevski2025}. These arise without invoking super-Eddington growth, offering a seeding route to part of a population usually explained by extreme accretion. \textsc{Aurora} does not, however, reproduce the full observed population: the highest mass ratios, which \citet{Chon} show to require super-Eddington growth in their \ac{lrd} systems, remain beyond its reach if the observed virial masses are taken at face value (\Cref{sec:pathway}). Those ratios are more likely to be biased high than low, through the selection of the observed sample \citep{Li2025} and, where the broad lines are not virial, through the virial mass estimates themselves (\citealt{Rusakov2026}; but see \citealt{Brazzini2025}).

\item Among seeded galaxies in haloes above $10^{12}\,\text{M}_\odot/h$ at $z \simeq 3.1$, the quiescent fraction rises from $0.40$ in the stock model to $0.48$ in the calibrated configuration and to $0.68$ at $\check{\alpha} = 2.6$. The enhancement weakens towards $z = 0$, as self-regulation erases the seed dependence of the black hole population. The signal is therefore strongest precisely at the epochs which the deep surveys are currently probing, and less apparent at low redshift, where models are conventionally calibrated. Reproducing the abundance of \acp{mqg}, however, requires highly resolved formation histories from early seeds, and the primordial environments of their progenitors, neither of which \textsc{Millennium} provides. We thus defer this objective to a subsequent paper, where \textsc{Aurora} will be applied to a simulation with superior mass and time resolution.

\item At $z \simeq 3.1$ the calibrated configuration raises the quiescent fraction in the densest tercile of seeding overdensity by $+0.10$ relative to the stock model and lowers it in the sparsest by $0.028$, constituting an environmental gradient of $\Delta\Theta = +0.13$ between them (\Cref{sec:quiescence}). Only the overdensity term, being centred on zero, can move the two ends in opposite directions, and no uniform change in the efficiency of feedback can reproduce the pattern. Rerunning the calibrated model with the environments scrambled, preserving the seed mass spectrum but destroying its correspondence with the density field, collapses the environmental gradient $\Delta\Theta$ from $+0.132$ to $-0.008$, and removes both the deficit in voids and the excess in nodes. An excess of early quiescent systems in overdense environments, accompanied by a deficit in voids, is therefore the most direct signature of environmental seeding in this work. Testing it observationally is a future endeavour, as the observational benchmarks used here do not include voids.

\item Each coupling steepens the environmental gradient $\Delta\Theta$ most strongly in its own variable (\Cref{sec:quiescence}, \cref{fig:fqenv_alpha,fig:fqenv_beta}). At $z \simeq 3.1$, raising $\check{\alpha}$ steepens the gradient across overdensity terciles from $+0.13$ to $+0.36$ while leaving the skew gradient unchanged, at $+0.072$. Raising $\check{\beta}$ steepens the skew gradient from $+0.072$ to $+0.16$, but also the overdensity gradient, from $+0.13$ to $+0.18$, so of the two only the overdensity coupling is neutral in the other's variable. The halo mass coupling leaves the quiescent fraction, and its enhancement in every tercile, within $0.007$ of the calibrated values, acting instead on the mass budget, where it raises $\eta$ by $14$ per cent. An observed dependence of the quiescent fraction on both variables would therefore constrain the two couplings separately, rather than yielding a single degenerate amplitude.

\item While the coupling strengths of the seeding law directly influence the scatter of the galaxy-black hole mass relation, producing some overmassive black holes (\Cref{sec:pathway}), the median galaxy statistics against which the base model is calibrated are left intact across the full sweep (\cref{fig:sweep_smbh}). The seeding prescription can therefore be varied freely, with the black hole population responding by factors of a few, without disturbing the calibration. Seeding can thus be explored as an open question rather than fixed as a precondition, which is what allows the sweeps presented here to be read as predictions rather than as recalibrations.
\end{itemize}

Taken together, these results establish that an environmentally graded seeding prescription leaves a real and measurable imprint on the galaxy population of massive haloes long after the seeding events themselves, surviving both self-regulated accretion and the degeneracy of the $z=0$ calibration (\Cref{sec:disc_signal}). The seeding prescription is also extensible; the equations governing the seed mass spectrum can be modified to include metallicity, irradiation, halo assembly rate, or any further quantity carried by the trees or the \acl{sam}, in the spirit of the channel-resolved seeding models of \citet{Spinoso2023} and \citet{Bhowmick2024}, but remaining applicable to gravity-only volumes (\Cref{sec:disc_extend}).

The wider implication concerns what can still be learned about the first black holes from the galaxies which now host them. Seeding is usually treated as a detail to be fixed before the interesting physics begins, on the reasoning that self-regulated accretion erases whatever the seed contributed, as the featureless high redshift black hole mass function appears to confirm \citep{Taylor2025}. The results presented here qualify that reasoning rather than contradict it: over $3.5 < z < 6$, where that mass function is measured, \textsc{Aurora}'s is at least as featureless as the stock model's, yet at $z \simeq 4.2$ the two visibly differ in quiescence across overdensity, by an environmental gradient of $\Delta\Theta = +0.18$ (\Cref{sec:disc_signal}). The seed of an individual black hole is indeed diluted almost completely by the time its host is massive, but the population is not, and the environmental grading imposed at seeding survives in a form which the mass function alone cannot show: in which galaxies quench, at which epochs, and in which surroundings. Information about the first black holes is therefore retained in places where it has not usually been sought.

That has a practical consequence, because those places are the ones which deep surveys are now mapping. Quiescent galaxies, \acp{lrd} and luminous quasars in the first billion years \citep{Fan2023, Glazebrook2024, Akins2025, Labbe, EuclidCollaboration2026} are precisely the rare, environmentally extreme systems on which an environment-dependent seeding prescription and an environment-blind one diverge most sharply, and they are being counted in volumes which only dark matter based models can populate. Whether the earliest massive black holes formed heavy in rare, deep potential wells, or light and grew rapidly where the gas was densest, is a question their descendants may yet answer. \textsc{Aurora} is one framework in which this question can be tested in high fidelity semi-analytic catalogues and in deep observations.

\section*{Acknowledgements}

This research was generously funded under Australian Research Council Laureate Fellowship Grant FL180100060 and NASA Grant JWST-ERS-2565. The authors value the \textsc{OzStar} Supercomputing Environment for institutional access to the simulation data and computational resources required to complete this work.

\section*{Data Availability}

Data and code are currently available upon reasonable request to the principal author.

\bibliographystyle{mnras}
\bibliography{biblio}

\bsp	
\label{lastpage}
\end{document}